\documentclass[12pt]{article}

\usepackage{amssymb}
\usepackage{color}
\usepackage{epsfig,amssymb,amsfonts,amsmath,graphicx,dsfont,cite,xfrac}
\usepackage{authblk}
\usepackage{subfigure}
\definecolor{mygray}{gray}{0.5}

\usepackage{cite}
\usepackage[colorlinks=true,linkcolor=blue,citecolor=red]{hyperref}

\newcommand{\be}{\begin{equation}}
\newcommand{\ee}{\end{equation}}
\newcommand{\bea}{\begin{eqnarray}}
\newcommand{\eea}{\end{eqnarray}}

\title{Bi-Orthogonal Approach to Non-Hermitian Hamiltonians with the Oscillator Spectrum: Generalized Coherent States for Nonlinear Algebras}

\author[${}$]{Oscar Rosas-Ortiz}
\author[${}$]{Kevin Zelaya}

\affil[${}$]{\footnotesize Physics Department, Cinvestav, AP 14-740, 07000
M\'exico City, Mexico}

\date{}
\begin{document}

\maketitle

\begin{abstract}
A set of Hamiltonians that are not self-adjoint but have the spectrum of the harmonic oscillator is studied. The eigenvectors of these operators and those of their Hermitian conjugates form a bi-orthogonal system that provides a mathematical procedure to satisfy the superposition principle. In this form the non-Hermitian oscillators can be studied in much the same way as in the Hermitian approaches. Two different nonlinear algebras generated by properly constructed ladder operators are found and the corresponding generalized coherent states are obtained. The non-Hermitian oscillators can be steered to the conventional one by the appropriate selection of parameters. In such limit, the generators of the nonlinear algebras converge to generalized ladder operators that would represent either intensity-dependent interactions or multi-photon processes if the oscillator is associated with single mode photon fields in nonlinear media.
\end{abstract}


\section{Introduction}

In ordinary quantum mechanics the dynamical variables ${\cal O}$ that are susceptible to measurement are called {\em observables}. These are usually represented by self-adjoint operators $O=O^{\dagger}$ whose eigenvectors form a complete set (i.e., the operators $O$ are {\em Hermitian}). The latter means that the superposition principle holds. In turn, the reason to restrict $O$ to be self-adjoint is merely practical since it is the simplest form to associate its eigenvalues with the result of a measurement of ${\cal O}$, which ``must always give a real number as result'' \cite{Dir58}. However, the reality of the results of any measurement  does not imply that the related observables must be represented by self-adjoint operators. Actually, there is a wide class of operators that have real spectrum although they are not self-adjoint. Notable examples are the ${\cal PT}$-symmetric Hamiltonians \cite{Ben99,Ben05}, the pseudo-Hermitian Hamiltonians \cite{Mos02,Mos10} and the non-Hermitian Hamiltonians generated by supersymmetry, see e.g. \cite{Can98,And99,Zno00,Bag01,Mil02,Ros03,Sin04,Ros15}. Such operators could also represent observables. In practice, they are useful for modeling systems whose phenomenology cannot be explained in terms of the conventional Hermitian approach \cite{Moi11,Bag16}. Some applications include the propagation of light in media with complex-valued refractive index \cite{Gom02}, transition probabilities in multi-photon processes \cite{Fai87} and the refinement of diverse techniques of measurement \cite{Sim17}.

Recently, we have introduced a class of one-dimensional Hamiltonians $H_{\lambda}$ whose eigenvalues are all real although the related potentials $V_{\lambda}$ are complex-valued functions \cite{Ros15}. That is, the operators $H_{\lambda}$ are not self-adjoint so that their eigenvectors are not mutually orthogonal. This implies that the Sturm--Liouville theory \cite{Inc56,Amr05}, which is useful to analyze the completeness of the eigenvectors of any self-adjoint operator, does not apply in the study of $H_{\lambda}$. Thus, it is difficult to determine whether any $H_{\lambda}$ satisfies the conditions for representing an observable or not. We may start by assuming that the measurement of a given dynamical variable will give as result any of the eigenvalues of $H_{\lambda}$. The main problem is to find a way to satisfy the superposition principle since the eigenvectors of $H_{\lambda}$ are not necessarily complete. Nevertheless, we have shown \cite{Jai17} that the real and imaginary parts of the eigenfunctions of $H_{\lambda}$ (i.e., the eigenvectors of $H_{\lambda}$ in position-representation) satisfy interlacing theorems that are very similar to those of the Hermitian approaches. The latter is a clear evidence that the eigenfunctions of $H_{\lambda}$ might be complete and is consistent with the results reported in \cite{Ben00} for ${\cal PT}$-symmetric Hamiltonians. However, the reality of the spectrum of $H_{\lambda}$ does not require the invariance under ${\cal PT}$ transformations since $H_{\lambda}$ is the result of applying the appropriate Darboux transformation on a given Hermitian Hamiltonian $H$. The main point is that the related complex-valued potentials $V_{\lambda}$ satisfy a condition of {\em zero total area} (the imaginary part of $V_{\lambda}$ is continuous in $\mathbb R$ and its integral over all the real line is equal to zero), which includes the ${\cal PT}$-symmetry as a particular case  \cite{Jai17}. 

The process of adding quantum states to give new quantum states is connected with a mathematical procedure that is always permissible \cite{Dir58}. This demands indeterminacy in the results of observations and was recognized as a breaking point from the classical ideas since the dawn of quantum theory \cite{Dir30}. To ensure that such a fundamental principle is satisfied by the states associated to $H_{\lambda}$, in this work we extend the orthonormal relation obeyed by the eigenvectors of $H$ to an orthonormal property which follows from the simultaneous consideration of the eigenvectors of $H_{\lambda}$ and those of $H_{\lambda}^{\dagger}$. The approach provides a mathematical structure for working with $H_{\lambda}$ and $H_{\lambda}^{\dagger}$ as  if they were two different faces of the same Hermitian Hamiltonian. Thus, the entire set of eigenvectors of $H_{\lambda}$ and $H_{\lambda}^{\dagger}$ form a {\em bi-orthogonal system} so that closure relations can be introduced to accomplish the superposition principle. Moreover, the algebraic properties of the operators that act on the eigenvectors of the non-Hermitian Hamiltonians are easily identified.

The physical model discussed in the present work is represented by a family of non-Hermitian operators $H_{\lambda}$ whose spectrum includes all the energies of the harmonic oscillator $E_n$, $n \geq 0$, plus an additional real eigenvalue $\epsilon < E_0$. Remarkably, the Hermitian oscillator-like systems reported in e.g. \cite{Abr80,Mie84}, as well as the conventional oscillator, are recovered as particular cases from our results. The non-Hermitian oscillators are constructed as Darboux transformations of the conventional oscillator and admit two different kinds of ladder operators which give rise to two different families of generalized coherent states. In this context we would like to emphasize that a coherent state is essentially a superposition of the energy eigenvectors of the harmonic oscillator \cite{Gla07,Kla68} (see also \cite{Kla60}). For systems other than the oscillator, the so-called generalized coherent states are concrete superpositions of the eigenvectors of a given observable \cite{Per86,Bar71}. Thus, to obtain superpositions with properties like those of the coherent states, it is useful to have at hand a complete set of eigenvectors. As the latter is not necessarily the case for non-Hermitian Hamiltonians, the construction of coherent states is a big challenge for such a class of operators in general. However, as we are going to see, the bi-orthogonal approach introduced in this work permits the derivation of such states in simple form. More specifically, we are going to deal with generalized coherent states that are bi-orthogonal superpositions of the energy eigenvectors of non-Hermitian oscillators. Within an approach that is very close to the one presented here, in \cite{Tri09,Bag11} and \cite{Bec01} are reported generalized coherent states for pseudo-boson systems and the  $\lambda$-deformed non-Hermitian oscillators, respectively. Other generalized coherent states have been obtained for ${\cal PT}$-symmetric oscillators either by modifying the normalization of states \cite{Bag01b} or by using the Gazeau-Klauder formalism \cite{Roy06}. 
 
On the other hand, the non-Hermitian oscillators studied in this work have the striking feature that the state belonging to the ground energy determines a class of ladder operators that satisfy a {\em quadratic polynomial Heisenberg algebra}. Indeed, such state is annihilated by both of the ladder operators. In turn, the state of the first excited energy is also canceled by the annihilation operator. The above profile leads to a variety of generalized coherent states that are eigenvectors of the annihilation operator belonging to complex eigenvalues $z$, with the eigenvalue $z=0$ twice degenerate. In addition, we find another algebra that does not depend on the ground energy of the system and is a common property of the non-Hermitian oscillators. This is called {\em distorted Heisenberg algebra} and is parameterized by a non-negative number $w \geq 0$. The generators are two additional ladder operators that also annihilate the ground energy eigenstate while the state of the first excited energy is still canceled by the annihilator. Thus, the space of states decomposes into the direct sum of subspaces where the Heisenberg algebra holds. The eigenvalue $z=0$ of the corresponding coherent states is also twice degenerate.

Each one of the above described sets of coherent states is an over-complete basis in the bi-orthogonal system of the non-Hermitian oscillators. They permit the construction of (i) a space of analytic entire functions (the {\em Fock-Bargmann representation} \cite{Foc28,Bar61}) for which the action of the ladder operators is expressed in terms of the multiplication by the eigenvalue $z$ and the derivatives with respect to $z$, and (ii) the matrix representation of the density operator in diagonal form (the {\em Glauber-Sudarshan $P$-representation} \cite{Gla63,Sud63}). Such a representation has a noticeable property: unlike the conventional oscillator, the eigenstate of the first excited energy belonging to the non-Hermitian oscillators is $P$-represented by a delta distribution.

The paper is organized as follows. In Sec.~\ref{oscillators} the non-Hermitian oscillator-like Hamiltonians $H_{\lambda}$ and $H_{\lambda}^{\dagger}$ are derived. It is shown that they are intertwined with the initial Hamiltonian $H$ by a pair of operators that also factorize $H$. The bi-orthogonal system formed by the eigenvectors of $H_{\lambda}$ and $H_{\lambda}^{\dagger}$ is constructed and the basic rules for operating with are introduced. In Sec.~\ref{transfT} we analyze the transformations of vectors and operators that are associated with the intertwining relationships. The Sec.~\ref{algebras} is devoted to the analysis of the algebras that are satisfied by the ladder operators of the non-Hermitian oscillators. The corresponding generalized coherent states are derived in Sec.~\ref{gcs} and the construction of the Fock-Bargmann spaces as well as the $P$-representation of operators is developed in Sec.~\ref{continuo}. We consider the special case of non-Hermitian oscillators with equidistant spectrum in Sec.~\ref{equidistant}. Besides, we show that some results for Hermitian oscillator-like Hamiltonians already reported in the literature can be recovered from ours as particular cases. Some applications involving models of multi-photon processes are discussed. Finally, in Sec.~\ref{conclu} we give some concluding remarks.

\section{Oscillator-like Hamiltonians}
\label{oscillators}

Along this work the conventional expression for the dimensionless Hamiltonian of  the harmonic oscillator $H_{osc}= \frac12 \left( \hat p^2 + \hat x^2 \right)$ will be replaced by its `mathematical' form 
\be
H = \hat p^ 2 + \hat x^ 2.
\label{ham1a}
\ee
The latter because the factor $\sfrac12$ (which is commonly preferred in physics) produces unnecessary complications to express in simple form the formulae we are going to deal with. Then, the traditional expression of the energy spectrum $n+\frac12$ will be replaced by $E_n= 2n+1$, with $n=0,1,2,\ldots$ The `mathematical' notation does not affect the relationships between the position $\hat x$ and momentum $\hat p$ operators, their commutation rule and uncertainty inequality are as usual $[\hat x, \hat p]=i \mathbb I$, $\Delta \hat x \Delta \hat p \geq \tfrac12$,
where $\mathbb I$ stands for the identity operator\footnote{Hereafter the product between any complex-valued function $f$ and the identity operator $\mathbb I$ will be abbreviated to just $f$.}. Notice however the modification (by a factor 2) in the commutators $[H,\hat x]= -2i \hat p$ and $[H, \hat p]=2i \hat x$. The Hamiltonian (\ref{ham1a}) can be factorized as $H= \hat a^{\dagger} \hat a +  1$, where the operators $\hat a^{\dagger}$ and $\hat a$ satisfy the commutator relation $[\hat a , \hat a^{\dagger} ] =2$. Introducing the number operator $\hat N =\frac12 \hat a^{\dagger} \hat a$ we also have $H= 2\hat N +1$, with $\hat N \vert n \rangle = n \vert n \rangle$. As in the ordinary case, the position-representation of the {\em number eigenvectors} $\vert n \rangle$ is given by the Hermite polynomials. In contrast, the action of the ladder operators on these vectors includes a factor $\sqrt 2$. Namely, $\hat a \vert n \rangle = \sqrt{2n} \vert n-1 \rangle$ and $\hat a^{\dagger} \vert n \rangle= \sqrt{2(n+1)} \vert n+1 \rangle$, with $n\geq 0$. In position-representation, $\hat x = x$, $\hat p = -i\frac{d}{dx}$, the ladder operators and the Hamiltonian acquire the differential form $\hat a = \frac{d}{dx} + x$, $\hat a^{\dagger} = -\frac{d}{dx} + x$, and $H= -\frac{d^2}{dx^2} +x^2$.

\subsection{Generalized factorization}

The factorization of $H$ is not restricted to the above algebraic expressions. In fact, we may look for a new pair of operators $A$, $B$, and a real constant $\epsilon$ such that
\be
H=AB +\epsilon.
\label{ham3a}
\ee
The self-adjointness of $H$ requires
\be
H =B^{\dagger} A^{\dagger} +\epsilon.
\label{ham3b}
\ee
One can show that $A=B^{\dagger}$ is a sufficient but not a necessary condition to solve (\ref{ham3a})-(\ref{ham3b}) in general \cite{Ros15}. For if we define the position-representation of the {\em factorizing operators} as
\be
A= -\frac{d}{dx} + \beta(x), \quad B= \frac{d}{dx} + \beta(x),
\label{factors}
\ee
with $\beta(x)$ a complex-valued function, then the adjoint expressions are
\be
A^{\dagger}= \frac{d}{dx} + \beta^*(x), \quad B^{\dagger}= -\frac{d}{dx} + \beta^*(x),
\label{factors2}
\ee
where the symbol ${}^*$ represents complex-conjugation. Clearly $A\neq B^{\dagger}$ and $A^{\dagger} \neq B$ whenever $\mbox{Im} \beta (x) \neq 0$. After introducing (\ref{factors}) in (\ref{ham3a}) we obtain the Riccati equation
\be
-\beta'(x) + \beta^2(x) = x^ 2-\epsilon.
\label{rica1}
\ee
If (\ref{factors2}) is now substituted into (\ref{ham3b}) the result gives the complex-conjugate of (\ref{rica1}), so that the self-adjointness of $H$ is automatically satisfied. The general form of $\beta(x)$ can be written as \cite{Ros15}
\be
\beta(x)= -\frac{\alpha'(x)}{\alpha(x)} +i \frac{\lambda}{\alpha^2(x)}, \quad \lambda \in \mathbb R,
\label{beta}
\ee
where $\alpha(x)$ is the real-valued function 
\be
\alpha(x) = \left[ a u^2_1(x) + b u_1(x) u_2(x)+ c u_2^2(x) \right]^{1/2},
\label{alpha1}
\ee
with $\{ a,b, c\}$ a set of non-negative parameters such that $4ac-b^2 = 4 \lambda^2$, and
\be
u_1(x)=  {}_1F_1 \left( \frac{1-\epsilon}{4}, \frac12;  x^2 \right) e^{-x^2/2}, \quad  
u_2(x)=  {}_1F_1 \left( \frac{3-\epsilon}{4}, \frac32;  x^2 \right) x e^{-x^2/2}. 
\label{alpha2}
\ee
The expression ${}_1F_1(a,c,z)$ stands for the confluent hypergeometric function \cite{Olv10}. 

\subsubsection{Non-Hermitian Hamiltonians}
\label{211}

Reversing the order of $A$ and $B$ in the product (\ref{ham3a}) we arrive at the operator
\be
BA = -\frac{d^2}{dx^2} + \beta'(x) + \beta^2(x).
\label{product}
\ee
The introduction of a function $V_{\lambda}(x)$ such that
\be
\beta'(x) + \beta^2(x) + \epsilon= V_{\lambda}(x),
\label{rica2}
\ee
leads to the Schr\" odinger operator 
\be
H_{\lambda}=BA +\epsilon = -\frac{d^2}{dx^2} + V_{\lambda} (x).
\label{ham5}
\ee
The subtraction of (\ref{rica1}) from (\ref{rica2}) shows that $V_{\lambda}(x)$ is a Darboux transformation of the oscillator potential
\be
V_{\lambda} (x) =x^ 2 + 2\beta'(x).
\label{pot2a}
\ee
Hereafter, to avoid singularities in the real-valued functions
\be
\mbox{Re} V_{\lambda}(x) = 2 \epsilon - x^2 + 2  \frac{\alpha'(x)}{\alpha(x)}, 
\qquad 
\mbox{Im} V_{\lambda}(x) = -4\lambda \frac{ \alpha'(x) }{ \alpha^3(x)},
\label{potparts}
\ee
we shall use functions $\alpha(x)$ with no zeros in $\mathbb R$. 

The interlacing properties of the eigenfunctions of $H_{\lambda}$ are not easily guessed for arbitrary complex-valued potentials $V_{\lambda}(x)$ \cite{Jai17}. However, the {\em condition of zero total area}
\be
\int_{\mathbb R} \mbox{Im} V_{\lambda}(x)  dx= \left. \frac{2\lambda}{\alpha^2(x)} \right\vert_{\mathbb R} =0
\label{zero}
\ee
implies that the related probability densities are such that (i) the number of their maxima increases as the level of the energy (ii) the distribution of such maxima is quite similar to that of the Hermitian problems, and (iii) the points of zero probability that are usual in the Hermitian problems are substituted by local minima (see Conjecture~5.1 in \cite{Jai17}). Moreover, given $\lambda \neq 0$, the simplest form to satisfy (\ref{zero}) is by constructing $\alpha(x)$ such that $\vert \alpha(x) \vert \rightarrow +\infty$ as $\vert x \vert \rightarrow +\infty$. This last produces $\mbox{Im} V_{\lambda}(x)/\mbox{Re} V_{\lambda}(x) \rightarrow 0$ as $\vert x \vert \rightarrow +\infty$. Then, it can be shown that there are no degenerate eigenvalues of $H_{\lambda}$ and that the corresponding eigenfunctions are normalizable on $\mathbb R$ (see, e.g. Sec.~5 of \cite{And07}).

\begin{figure}[htb]

\centering
\subfigure[$\epsilon= 0.5$]{\includegraphics[width=0.3\textwidth]{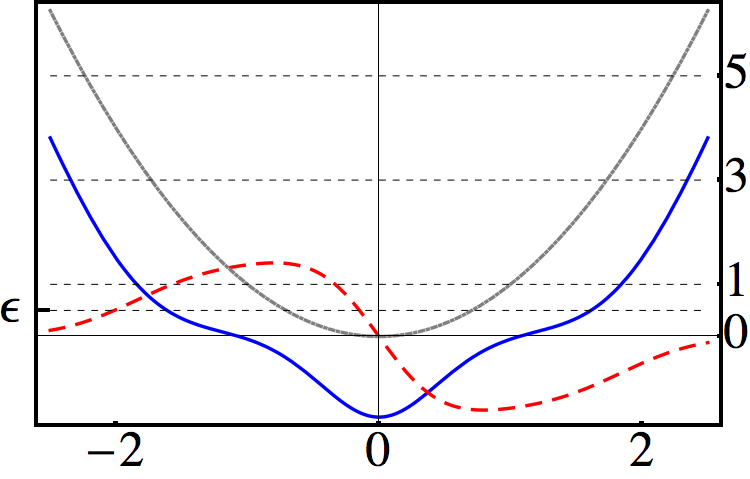}} 
\hspace{1ex}
\subfigure[$\epsilon= -3$]{\includegraphics[width=0.3\textwidth]{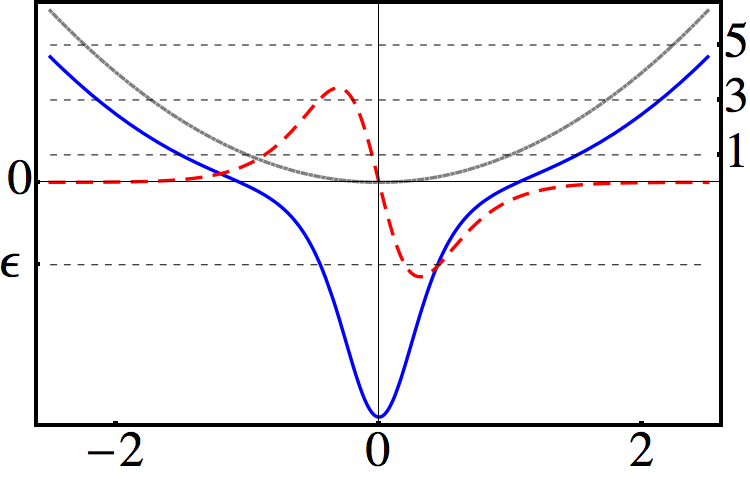}}
\hspace{1ex}
\centering
\subfigure[$\epsilon= -5$]{\includegraphics[width=0.3\textwidth]{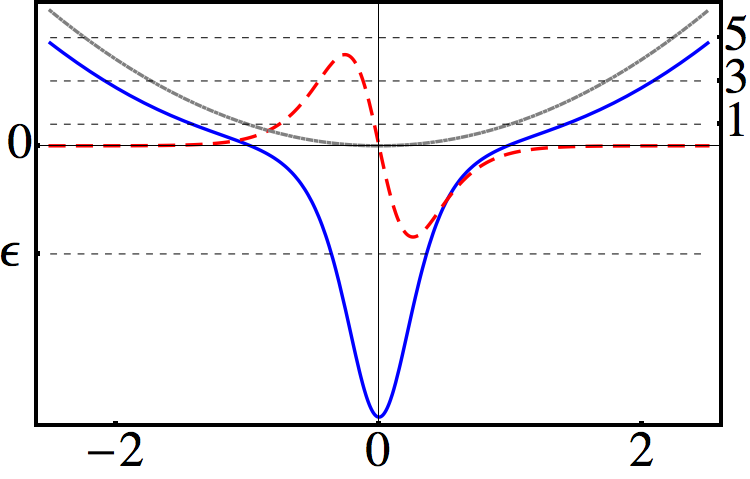}}

\centering
\subfigure[$\epsilon= 0.5$]{\includegraphics[width=0.3\textwidth]{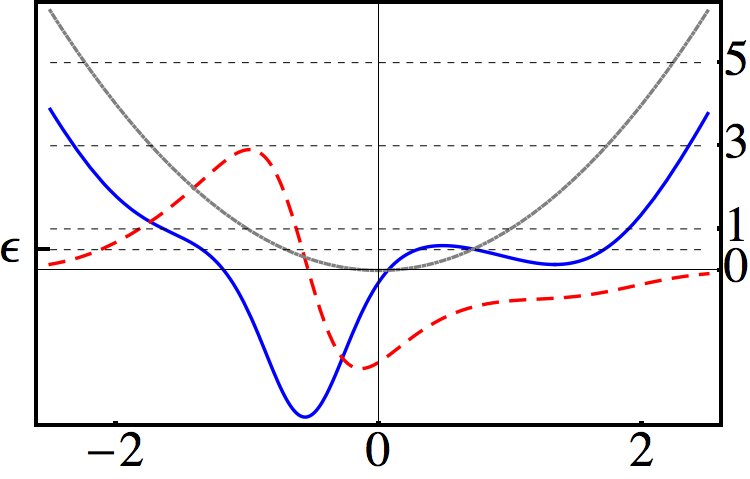}} 
\hspace{1ex}
\subfigure[$\epsilon= -3$]{\includegraphics[width=0.3\textwidth]{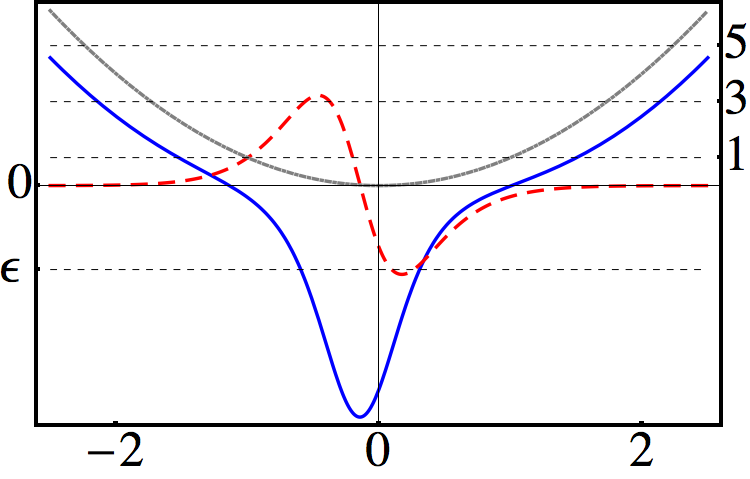}}
\hspace{1ex}
\centering
\subfigure[$\epsilon= -5$]{\includegraphics[width=0.3\textwidth]{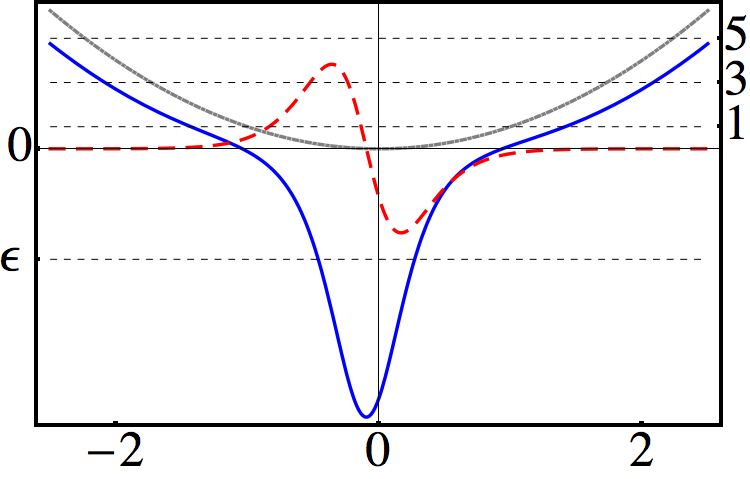}}

\caption{\footnotesize 
(Color online) The real (solid-blue) and imaginary (dashed-red) parts of the complex-valued potential $V_{\lambda}(x)$ defined in (\ref{pot2a}) for the indicated ground energies $\epsilon <1$. In all the cases the horizontal dotted lines represent allowed energies, and the oscillator potential (dotted-gray) is included as a reference. 
}
\label{f_pot}
\end{figure}

The panel shown in Fig.~\ref{f_pot} includes a series of complex-valued oscillator-like potentials  $V_{\lambda}(x)$ that satisfy the condition of zero total area (\ref{zero}). Those in the upper row are ${\cal PT}$-symmetric while the ones in the lower row are non-invariant under the ${\cal PT}$ reflection. We have selected cases for which the gap of the lowest two energies is different from the gap of any other pair of consecutive energies, so the spectrum is not strictly equidistant. For the determination of the energy eigenvalues see Sec.~\ref{solution}; the equidistant case will be discussed in detail in Sec.~\ref{equidistant}. 

Clearly, for $\lambda \neq 0$, the Schr\"odinger operator $H_{\lambda}$ is non-Hermitian since it is formally different from its adjoint
\be
H_{\lambda}^{\dagger}= A^{\dagger} B^{\dagger}+\epsilon = -\frac{d^2}{dx^2} + V_{\lambda}^* (x).
\label{ham6}
\ee
As a consequence, the completeness of the eigenvectors of $H_{\lambda}$ is not granted a priori  because they are not mutually orthogonal. The same can be said about the eigenvectors of $H_{\lambda}^{\dagger}$. This is an usual difficulty in the spectral problem of non-Hermitian Schr\"odinger operators and implies that the superposition of eigenvectors is not able in the ordinary form. However, the Hamiltonians $H_{\lambda}$ and $H_{\lambda}^{\dagger}$ belong to the class of {\em softly non-Hermitian} Hamiltonians introduced in \cite{And07} as they certainly do not involve complex coordinates in the definition of asymptotic boundary conditions. Thus, a bi-orthogonal resolution of the identity is viable since there is no more than a Jordan cell made of a normalizable eigenfunction of $H_{\lambda}$ and the corresponding one of $H_{\lambda}^{\dagger}$ for any of the energy eigenvalues (see Sec.~\ref{31} for details and Ref.~\cite{And07}). That is, our non-Hermitian model is free of the puzzles with self-orthogonal states described in \cite{Sok06}.

\subsubsection{Solution to the eigenvalue problem}
\label{solution}

From (\ref{ham3a}) and (\ref{ham5}) we obtain the intertwining relationship
\be
H_{\lambda} B= BH, \qquad HA = AH_{\lambda},
\label{inter1}
\ee
so that $B\vert n \rangle$ is  eigenvector of $H_{\lambda}$ with eigenvalue $E_n$. In addition, the solution of $A\vert \psi_{\epsilon} \rangle =0$ must be considered since it also satisfies $(H_{\lambda} - \epsilon) \vert \psi_{\epsilon} \rangle =0$. Then, we may write 
\be
H_{\lambda} \vert \psi_n \rangle = E^{(\lambda)}_n \vert \psi_n \rangle, \quad n\geq 0,
\label{eigen1}
\ee
with
\be
\vert \psi_{n+1} \rangle = \theta_{n+1} B \vert n \rangle, \quad \vert \psi_0 \rangle = \theta_0  \vert \psi_{\epsilon} \rangle, \quad E^{(\lambda)}_{n+1} =E_n, \quad E^{(\lambda)}_0 =\epsilon.
\label{kets1}
\ee
The $\theta_n$ in (\ref{kets1}) stand for the normalization constants, which are to be fixed. Now, let us look at the eigenvalue equation 
\be
\overline H_{\lambda} \vert \overline \psi_n \rangle = \overline E^{(\lambda)}_n \vert \overline \psi_n \rangle, \quad n\geq 0,
\label{eigen3}
\ee
where $\overline H_{\lambda} \equiv H_{\lambda}^{\dagger}$. As with the previous case, the spectral properties of $\overline H_{\lambda}$ are intertwined with those of $H$. In fact, (\ref{ham3b}) and (\ref{ham6}) lead to the relationship
\be
\overline H_{\lambda} A^{\dagger} = A^{\dagger} H, \qquad HB^{\dagger} =B^{\dagger} \overline H_{\lambda},
\label{inter2}
\ee
so that 
\be
\vert \overline \psi_{n+1} \rangle = \overline \theta_{n+1} A^{\dagger} \vert n \rangle, \quad \vert \overline \psi_0 \rangle = \overline \theta_0 \vert \overline \psi_{\epsilon} \rangle, \quad \overline E^{(\lambda)}_{n} = E^{(\lambda)}_n, \quad n\geq 0,
\label{kets2}
\ee
where $\vert \overline \psi_{\epsilon} \rangle$ is solution of $B^{\dagger} \vert \overline \psi_{\epsilon} \rangle =0$ as well as eigenvector of $\overline H_{\lambda}$ with eigenvalue $\overline E^{(\lambda)}_0=\epsilon$. 

\begin{figure}[htb]
\centering\includegraphics[width=0.4\textwidth]{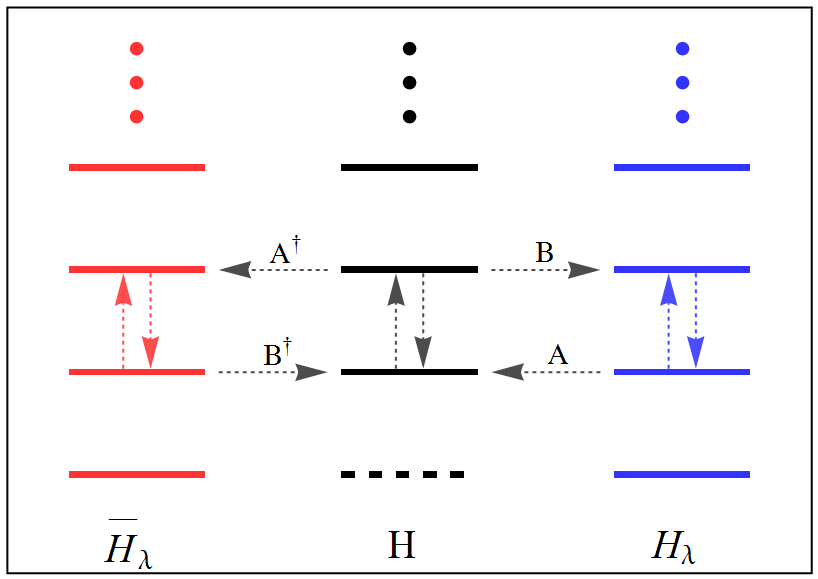}

\caption{\footnotesize 
(Color online) Energy spectrum of the oscillator Hamiltonian $H$ (black, center), and the non-Hermitian Schr\"odinger operators $H_{\lambda}$ (blue, right) and $\overline H_{\lambda}$ (red, left). The horizontal solid lines represent allowed energies in all the cases. 
The energy level appearing at the bottom of each one of the three spectra represents the eigenvalue $\epsilon$, which is forbidden for the oscillator (horizontal dashed line) since the corresponding eigenvector is not normalizable, compare with Fig.~\ref{f_pot}. As indicated by the horizontal arrows, the pairs of operators $A$, $B$, and $A^{\dagger}$, $B^{\dagger}$, produce the intertwining relationships (\ref{inter1}) and (\ref{inter2}) respectively. The vertical arrows represent the action of ladder operators in the respective set of eigenvectors. 
}
\label{f_spectra}
\end{figure}

After changing $n$ to $m$, the adjoint of Eqs.~(\ref{eigen1}) and (\ref{eigen3}) are respectively
\be
\langle \psi_m \vert  \overline H_{\lambda} = E^{(\lambda)}_m  \langle  \psi_m \vert, \quad m\geq 0
\label{eigen2}
\ee
and
\be
\langle \overline \psi_m \vert   H_{\lambda} = \overline E^{(\lambda)}_m \langle  \overline \psi_m \vert, \quad m\geq 0.
\label{eigen4}
\ee
We may now calculate the action of $\langle \overline \psi_m \vert$ on the right of (\ref{eigen1}), and the action of $\vert \psi_n \rangle$ on the left of (\ref{eigen4}). The subtraction of these results yields
\be
\left(E^{(\lambda)}_n -E^{(\lambda)}_m \right) \langle \overline \psi_m \vert \psi_n \rangle=0.
\label{cond1}
\ee
The similar operation with (\ref{eigen3}) and (\ref{eigen2}) gives
\be
\left(E^{(\lambda)}_m -E^{(\lambda)}_n \right) \langle \psi_m \vert \overline \psi_n \rangle=0.
\label{cond2}
\ee
Let us concentrate in Eq.~(\ref{cond1}). If $n\neq m$ one has $\langle \overline \psi_m \vert \psi_n \rangle = 0$ by necessity. The solution for $n=m$ is calculated by using (\ref{kets2}), (\ref{kets1}) and (\ref{ham3a}); one arrives at the complex numbers
\be
\langle \overline \psi_{n+1} \vert \psi_{n+1} \rangle = \overline \theta _{n+1}^{\, *} \theta_{n+1} \left(E_n -\epsilon \right), \qquad \langle \overline \psi_0 \vert \psi_0 \rangle = \overline \theta _0^{\, *} \theta_0  \, \langle \overline \psi_{\epsilon} \vert \psi_{\epsilon} \rangle.
\ee
Taking $\epsilon < E_0$, the real constants $\overline \theta_{n+1} =\theta_{n+1} = (E_n -\epsilon)^{-1/2}$ are sufficient to normalize the expression $\langle \overline \psi_{m+1} \vert \psi_{n+1} \rangle$. The case $\langle \overline \psi_0 \vert \psi_0 \rangle$ requires some caution, for if we assume that $ \langle \overline \psi_{\epsilon} \vert \psi_{\epsilon} \rangle$ is a complex number, then its modulus must be finite and different from zero. In such a case we use the polar form $\overline \theta_0^{\, *} = \theta_0 =\vert \langle \overline \psi_{\epsilon} \vert \psi_{\epsilon} \rangle \vert^{-1/2} e^{-i \chi/2}$. Then $\langle \overline \psi_m \vert \psi_n \rangle = \delta_{n,m}$, and the normalized solutions of the eigenvalue problem (\ref{eigen1}) are given by the set
\be
\begin{aligned}
\vert \psi_{n+1} \rangle= \frac{1}{\sqrt {E_n-\epsilon}} B \vert n \rangle, \quad E^{(\lambda)}_{n+1} = E_n \quad n\geq0,\\[1ex]
\vert \psi_0 \rangle = 
\vert \langle \overline \psi_{\epsilon} \vert \psi_{\epsilon} \rangle \vert^{-\frac12} \, e^{-i \frac{\chi}{2}}
\, \vert \psi_{\epsilon} \rangle, \quad E^{(\lambda)}_0 = \epsilon.
\end{aligned}
\label{sol1}
\ee
Using a similar procedure, from (\ref{cond2}) we obtain $\langle \psi_m \vert \overline \psi_n \rangle =\delta_{m,n}$, so that the normalized solutions of (\ref{eigen3}) are written as
\be
\begin{aligned}
\vert \overline \psi_{n+1} \rangle= \frac{1}{\sqrt {E_n-\epsilon}} A^{\dagger} \vert n \rangle, \quad \overline E^{(\lambda)}_{n+1} = E_n \quad n\geq0,\\[1ex]
\vert \overline\psi_0 \rangle =
\vert \langle \overline \psi_{\epsilon} \vert \psi_{\epsilon} \rangle \vert^{-\frac12} \, e^{-i \frac{\chi}{2}}
 \vert \overline\psi_{\epsilon} \rangle, \quad \overline E^{(\lambda)}_0 = \epsilon.
\end{aligned}
\label{sol2}
\ee
In the diagram of Fig.~\ref{f_spectra} we represent the form in which the spectra of $H$, $H_{\lambda}$ and $\overline H_{\lambda}$ are interrelated. Combining this information with that of Fig.~\ref{f_pot} we have a versatile picture of the spectral properties of the non-Hermitian Hamiltonians $H_{\lambda}$ and $\overline H_{\lambda}$.

\subsection{Bi-orthogonal system}

As indicated above, the generalized orthonormal relations
\be
 \langle \overline \psi_m \vert \psi_n \rangle = \langle \psi_m \vert \overline \psi_n \rangle =\delta_{m,n}, \quad n,m\geq 0,
\label{ortho1}
\ee
involve the eigenvectors of  $H_{\lambda}$ as well as those of $\overline H_{\lambda}$. The vector superpositions
\be
\vert \xi_{\lambda} \rangle =  \sum_{n=0}^{+\infty} c_n \vert \psi_n \rangle, \qquad 
\vert \overline \eta_{\lambda} \rangle = \sum_{n=0}^{+\infty} \overline b_n^{\, *} \vert \overline \psi_n \rangle, 
\label{super1}
\ee
are now permitted since (\ref{ortho1})  defines the Fourier coefficients $c_n = \langle \overline \psi_n \vert \xi_{\lambda} \rangle$ and $\overline b_n^{\, *} = \langle  \psi_n  
\vert \overline \eta_{\lambda} \rangle$. Besides, the notion of `orthogonality' between vectors arises by introducing the {\em bi-product} between $\vert \overline \eta_{\lambda} \rangle$ and $\vert \xi_{\lambda} \rangle$ as follows
\be
\langle \overline \eta_{\lambda} \vert \xi_{\lambda} \rangle=\sum_{n=0}^{+\infty} \overline b_n c_n = \langle \xi_{\lambda} \vert \overline \eta_{\lambda} \rangle^*.
\label{prod1}
\ee
Then, the vectors $\vert \overline \eta_{\lambda} \rangle$ and $\vert \xi_{\lambda} \rangle$ are {\em bi-orthogonal} if $\langle \overline \eta_{\lambda} \vert \xi_{\lambda} \rangle=0$. Now, let us assume that given $\vert \xi_{\lambda} \rangle$ there exists a vector $\vert \overline \xi_{\lambda} \rangle$ such that $\langle \overline \xi_{\lambda} \vert \xi_{\lambda} \rangle \geq0$. From  (\ref{super1}) we arrive at  the complex series
\be
\langle \overline \xi_{\lambda} \vert \xi_{\lambda} \rangle = \sum_{n=0}^{+\infty} c_n \langle \overline \xi_{\lambda} \vert \psi_n \rangle.
\label{prod2}
\ee
The latter becomes a sum of non-negative numbers by taking $\langle \overline \xi_{\lambda}  \vert \psi_n \rangle=c_n^*$. Then, the non-negative quantity
\be
\vert \vert \xi_{\lambda} \vert \vert^ 2= \langle \overline \xi_{\lambda} \vert \xi_{\lambda} \rangle = \sum_{n=0}^{+\infty} \vert c_n \vert^2 \geq0
\label{prod3}
\ee
is useful to introduce the {\em bi-norm} $\vert \vert \xi_{\lambda} \vert \vert$
 of the vector $\vert \xi_{\lambda} \rangle$. Of course, (\ref{prod3}) is also associated to the bi-norm of  $\vert \overline \xi_{\lambda} \rangle$ since $\langle \overline \xi_{\lambda} \vert \xi_{\lambda} \rangle= \langle \xi_{\lambda} \vert \overline \xi_{\lambda} \rangle$. For convergent series, the {\em normalization condition} is written as follows
\be
\vert \vert \xi_{\lambda} \vert \vert^ 2 = \sum_{n=0}^{+\infty} \vert c_n \vert^2= 1.
\label{normal1}
\ee
A pair of remarks are necessary: 

(i) The product (\ref{prod3}) implies that the {\em concomitant} $\vert \overline \xi_{\lambda} \rangle$ of $\vert \xi_{\lambda} \rangle$ must be expressed as
\be
\vert \overline \xi_{\lambda} \rangle = \sum_{n=0}^{+\infty} c_n \vert \overline \psi_n \rangle, \quad c_n = \langle \psi_n \vert \overline \xi_{\lambda} \rangle.
\label{super1b}
\ee
That is, $\vert \xi_{\lambda} \rangle$ and $\vert \overline \xi_{\lambda} \rangle$ share the same coefficients $c_n = \langle \overline \psi_n \vert \xi_{\lambda} \rangle= \langle \psi_n \vert \overline \xi_{\lambda} \rangle$ when they are expanded in their respective sets of eigenvectors 

(ii) As $\vert \psi_n \rangle$ and $\vert \overline \psi_n \rangle$ are concomitant of bi-norm $\vert \vert \psi_n \vert \vert = \vert \vert \overline \psi_n \vert \vert =1$, and the set $\{ \vert \psi_n \rangle \}_{n\geq 0}$ is bi-orthogonal to $\{ \vert \overline \psi_m \rangle \}_{m\geq 0}$, we would take the entire set $\{ \vert \overline \psi_m \rangle , \vert \psi_n \rangle \}_{m,n\geq 0}$ as the platform to construct the representation spaces of the non-Hermitian Schr\"odinger operators $H_{\lambda}$ and $\overline H_{\lambda}$.

The missing point to solve is the unicity of the superpositions (\ref{super1}). With this aim, let us look for a normalized eigenvector $\vert \overline \Phi_{\lambda} \rangle$ of $\overline H_{\lambda}$ which is bi-orthogonal to all the set $\{ \vert \psi_n \rangle \}_{n\geq 0}$. If by chance there exist such a vector, then it must be added to the set $\{ \vert \overline \psi_m \rangle \}_{m\geq 0}$. A similar logic operates for a possible missing state of $H_{\lambda}$. For $n\neq 0$, a simple calculation shows that $\vert \overline \Phi_{\lambda} \rangle$ must be a solution of $B^{\dagger} \vert \overline \Phi_{\lambda} \rangle=0$. However, the latter has been already considered; this is the ground-energy eigenvector $\vert \overline \psi_0 \rangle$, which is bi-orthogonal to all the $\vert \psi_{n+1} \rangle$. On the other hand, for $n=0$, the orthonormal relation (\ref{ortho1}) shows that any superposition of the vectors $\vert \overline \psi_{m+1} \rangle$ is bi-orthogonal to $\vert \psi_0 \rangle$. However, these superpositions are not eigenvectors of $H_{\lambda}$ in general. Only taking the simplest case $\vert \overline \Phi_{\lambda} \rangle = \vert \overline \psi_{m+1} \rangle$ we obtain the appropriate solution. Therefore, the set $\{ \vert  \overline \psi_m \rangle \}_{m \geq 0}$ includes all the eigenvectors of $\overline H_{\lambda}$ that are bi-orthogonal to any $\vert \psi_n \rangle$, with $n\neq m$. Equivalently, the set $\{ \vert   \psi_n \rangle \}_{n\geq 0}$ includes all the eigenvectors of $H_{\lambda}$ that are bi-orthogonal to any $\vert \overline\psi_m \rangle$. As a conclusion, the vector decompositions (\ref{super1}) are unique. This result is very important because it permits the introduction of the closure relations
\be
\mathbb I_{\lambda} = \sum_{n=0}^{+\infty} \vert \psi_n \rangle \langle \overline \psi_n \vert, \qquad \overline{\mathbb I}_{\lambda} = \sum_{n=0}^{+\infty} \vert \overline \psi_n \rangle \langle \psi_n \vert.
\label{iden3}
\ee
The vector decompositions (\ref{super1}) are now obtained by the action of $\mathbb I_{\lambda}$ and $\overline{\mathbb I}_{\lambda}$ on $\vert \xi_{\lambda} \rangle$ and $\vert \overline \eta_{\lambda} \rangle$, respectively. Now, in position-representation, the closure relations (\ref{iden3}) imply
\be
\delta (x-x')= \langle x \vert x' \rangle = \sum_{n=0}^{+\infty} \psi_n(x) \overline\psi_n^{\, *}(x') = \sum_{n=0}^{+\infty} \psi_n(x) \psi_n (x'),
\label{bibo1}
\ee
where we have used that $\vert \psi_n \rangle$ and $\vert \overline \psi_n \rangle$ are concomitant, so that $\overline \psi_n^{\, *}(x) = \psi_n(x)$. On the other hand, from (\ref{ortho1}) we have
\be
\delta_{m,n} = \langle \overline \psi_m \vert \psi_n \rangle = \int_{\mathbb R} dx \,  \psi_n(x) \overline \psi_m^{\, *}(x) = \int_{\mathbb R} dx \,  \psi_n(x) \psi_m (x).
\label{bibo2}
\ee
As the bi-norms derived from (\ref{bibo2}) do not vanish, the resolution of identity (\ref{bibo1}) holds \cite{Sok06}.

The above properties provide a mathematical structure for the simultaneous consideration of the sets $\{ \vert \psi_n \rangle \}_{n\geq 0}$ and $\{ \vert \overline \psi_m \rangle \}_{m\geq 0}$ that is not present if we consider them on their own. The entire set $\{ \vert \overline \psi_m \rangle , \vert \psi_n \rangle \}_{m,n\geq 0}$ defines the {\em bi-orthogonal system} \cite{Cur07,Bro14} that we shall use to construct arbitrary states of the non-Hermitian oscillators.

\section{Transformation theory}
\label{transfT}

Of special interest, the expressions (\ref{iden3}) open the possibility of working with the operators $H_{\lambda}$ and $\overline H_{\lambda}$ as if they were two different faces of a Hermitian Hamiltonian. Let ${\cal H}_{\lambda}$ be the set of all the vectors  that can be constructed as a bi-orthogonal superposition of the $\vert \psi_n \rangle$. Then, $\mathbb I_{\lambda}$ is an automorphism of ${\cal H}_{\lambda}$ which works as the identity operator. We write $\mathbb I_{\lambda} \in \mbox{Aut} \left( {\cal H}_{\lambda} \right)$, with $\mbox{Aut} \left( {\cal H}_{\lambda} \right)$ the set of automorphisms of ${\cal H}_{\lambda}$. As the vectors $\vert \psi_n \rangle$ have been constructed by the action of $B$ on the number eigenvectors $\vert n \rangle$, we say that ${\cal H}_{\lambda}$ is the representation space of $H_{\lambda}$ induced by $B$. Thus, the operator $B$ defines a mapping ${\cal H} \rightarrow {\cal H}_{\lambda}$ that connects the Hilbert space ${\cal H}$ of the number eigenvectors with the space of states of $H_{\lambda}$. The operator $A$ reverses the action of $B$ and maps the ground state $\vert \psi_0 \rangle$ of $H_{\lambda}$ into the null vector $\vert \varnothing \rangle$ of ${\cal H}$, see Fig.~\ref{f_spectra}. Therefore, the algebraic relationships (\ref{inter1}) imply the mappings
\be
B: {\cal H} \rightarrow {\cal H}_{\lambda} \quad \mbox{and} \quad A: {\cal H}_{\lambda} \rightarrow {\cal H}.
\label{map1}
\ee 
For completeness, let us write explicitly the mappings (\ref{map1}) of the corresponding vector basis
\be
B \vert n \rangle = \sqrt{E_n -\epsilon} \, \vert \psi_{n+1} \rangle, \quad A \vert \psi_{n+1} \rangle = \sqrt{E_n -\epsilon} \, \vert n \rangle, \quad A \vert \psi_0 \rangle = \vert \varnothing \rangle, \quad n\geq 0.
\label{bases1}
\ee
In similar form, denoting by $\overline {\cal H}_{\lambda}$ the set of all the vectors that can be expressed as a superposition of the $\vert \overline \psi_n \rangle$, one realizes that $\overline{\mathbb I}_{\lambda}$ is the identity operator in $\overline {\cal H}_{\lambda}$. Besides, we say that $\overline {\cal H}_{\lambda}$ is the representation space of $\overline H_{\lambda}$ induced by $A^{\dagger}$. Then, from (\ref{inter2}) one has a new pair of mappings
\be
A^{\dagger}: {\cal H} \rightarrow \overline {\cal H}_{\lambda} \quad \mbox{and} \quad B^{\dagger}: \overline{\cal H}_{\lambda} \rightarrow {\cal H},
\label{map2}
\ee 
where $B^{\dagger}$ reverses the action of $A^{\dagger}$ and maps the ground state $\vert \overline \psi_0 \rangle$ of $H_{\lambda}$ into the null vector $\vert \varnothing \rangle$ of ${\cal H}$, see Fig.~\ref{f_spectra}. The mappings (\ref{map2}) of the related vector basis are as follows
\be
A^{\dagger} \vert n \rangle = \sqrt{E_n -\epsilon} \, \vert \overline \psi_{n+1} \rangle, \quad B^{\dagger} \vert \overline \psi_{n+1} \rangle = \sqrt{E_n -\epsilon} \, \vert n \rangle, \quad B^{\dagger} \vert \overline \psi_0 \rangle = \vert \varnothing \rangle, \quad n\geq 0.
\label{bases2}
\ee
As immediate example consider an arbitrary (normalized) vector in ${\cal H}$,
\be
\vert \phi \rangle= \sum_{n=0}^{+\infty} a_n \vert n \rangle, \quad \sum_{n=0}^{+\infty}  \vert a_n \vert^2=1.
\label{tlin1}
\ee
This last is transformed into either
\be
\vert \phi^{(\lambda)} \rangle = c_{\phi} B \vert \phi \rangle = c_{\phi}  \sum_{n=0}^{+\infty} a_n \sqrt{E_n -\epsilon} \, \vert \psi_{n+1}  \rangle
\label{tlin2}
\ee
or
\be
\vert \overline \phi^{\, (\lambda)} \rangle = c_{\phi}  A^{\dagger} \vert \phi \rangle = c_{\phi}  \sum_{n=0}^{+\infty} a_n \sqrt{E_n -\epsilon} \, \vert \overline \psi_{n+1}  \rangle,
\label{tlin3}
\ee
with $c_{\phi} = \left( \langle H \rangle_{\phi}  -\epsilon \right)^{-1/2}$ the normalization constant and $\langle H \rangle_{\phi} \equiv \langle \phi \vert H \vert \phi \rangle$ the expectation value of $H$ calculated for $\vert \phi \rangle$. Notice that $\vert \phi^{(\lambda)} \rangle$ and $\vert \overline \phi^{\, (\lambda)} \rangle$ are concomitant. The usefulness of the above transformation will be clear in the sequel.

\subsection{Transformation of operators}
\label{31}

The identity operators introduced in the previous section are not the simplest automorphisms that can be constructed for the representation spaces ${\cal H}_{\lambda}$ and $\overline {\cal H}_{\lambda}$. Indeed, the ``dyadic'' operators
\be
X^{n,m} =\vert \psi_n \rangle \langle \overline \psi_m \vert, \qquad \overline X^{\, k,j} = \ \vert \overline \psi_k \rangle \langle \psi_j \vert, \qquad \left( X^{n,m}\right)^{\dagger}= \overline X^{\, m,n}, \qquad n,m,k,j \geq 0,
\label{hubbard}
\ee
are such that $X^{n,m} \vert \psi_r \rangle = \delta_{m,r} \vert \psi_n \rangle$ and $\overline X^{\, k,j} \vert \overline \psi_{\ell} \rangle = \delta_{j,\ell} \vert \overline \psi_k \rangle$. In general, for any $\vert \xi_{\lambda} \rangle \in {\cal H_{\lambda}}$ and $\vert \overline \eta_{\lambda} \rangle \in {\cal \overline H_{\lambda}}$ we have
\be
X^{n,m} \vert \xi_{\lambda} \rangle = c_m \vert \psi_n \rangle, \qquad \overline X^{\, k,j} \vert \overline \eta_{\lambda} \rangle =  \overline b_j^* \vert \overline \psi_k \rangle,
\label{hubbard2}
\ee
where $c_m$ and $\overline b_j^*$ are given in (\ref{super1}). The dyads (\ref{hubbard}) are known as {\em Hubbard operators} \cite{Enr13} and obey the multiplication rule $X^{n,m} X^{r,s}  = \delta_{m,r} X^{n,s}$,  $\overline X^{\, k,j}  \overline X^{\, i, \ell}= \delta_{j,i} \overline X^{\, k,\ell}$. In its simplest form, $X^{n,m}$ corresponds to an square matrix of infinite order for which all the entries are zero except the one at the $(n+1)$th row and the $(m+1)$th column, where it takes the value 1. A similar statement is true for $\overline X^{\, k,j}$. One of the most useful properties of these operators is that their diagonal representatives serve as `basis' for the closure relations introduced in (\ref{iden3}), namely
\be
\mathbb I_{\lambda}= \sum_{n=0}^{+\infty} X^{n,n}, \qquad \mathbb {\overline I}_{\lambda}= \sum_{k=0}^{+\infty} \overline X^{\, k,k}.
\label{Imatrix}
\ee
Thus, $\mathbb I_{\lambda}$ and $\mathbb {\overline I}_{\lambda}$ are diagonal matrices for which the non-zero entries are all equal to 1, as expected. An immediate application of these results is obtained by calculating the matrix-representation of $H_{\lambda}$ and $\overline H_{\lambda}$, it yields the same diagonal matrix
\be
H_{\lambda} = \sum_{n=0}^{+\infty} E_n^{(\lambda)} X^{n,n}, \qquad \overline H_{\lambda} = \sum_{k=0}^{+\infty} E_k^{(\lambda)} \overline X^{\, k,k}.
\label{Hmatrix}
\ee
The last expressions show that there is only one Jordan cell $X^{n,n}$, equivalently $\overline X^{\, n,n}$, made of the normalizable eigenvector $\vert \psi_n \rangle$ and its concomitant $\vert \overline \psi_n \rangle$ for each eigenvalue $E_n$, as it is indicated in \cite{And07} for the softly non-Hermitian Hamiltonians.

In general, the matrix-representation of any operators $O_{\lambda} \in  \mbox{Aut} \left( {\cal H}_{\lambda} \right)$ and $\overline O_{\lambda} \in  \mbox{Aut} \left( \overline {\cal H}_{\lambda} \right)$ is easily achieved
\be
O_{\lambda} = \sum_{n,m=0}^{+\infty} o_{n,m} X^{n,m}, \qquad 
\overline O_{\lambda} = \sum_{k,j=0}^{+\infty} \overline o_{k,j} \overline X^{\, k,j},
\label{matrix1}
\ee
where $o_{n,m} =\langle \overline \psi_n \vert O_{\lambda} \vert \psi_m \rangle$ and $\overline o_{k,j} =\langle  \psi_k \vert \overline O_{\lambda} \vert \overline \psi_j \rangle$. The {\em bi-orthogonal expectation values} are calculated as $\langle O_{\lambda} \rangle:= \langle \overline \xi_{\lambda} \vert O_{\lambda} \vert \xi_{\lambda} \rangle$ and $\langle \overline O_{\lambda} \rangle:= \langle  \xi_{\lambda} \vert \overline O_{\lambda} \vert \overline \xi_{\lambda} \rangle$. In particular, from (\ref{Imatrix}) one has $\langle \mathbb I_{\lambda} \rangle= \langle \mathbb {\overline I}_{\lambda} \rangle=1$, as expected. Besides, from (\ref{Hmatrix}), it is easy to verify that the non-Hermitian Schr\"odinger operators $H_{\lambda}$ and $\overline H_{\lambda}$ share the same expectation value $\langle H_{\lambda} \rangle= \langle \overline H_{\lambda} \rangle$.

Now, we can investigate the form in which the operator $O \in \mbox{Aut}({\cal H})$ is mapped into either $\mbox{Aut}({\cal H}_{\lambda})$  or $\mbox{Aut}(\overline {\cal H}_{\lambda})$. Let us write $O$ in Hubbard representation (to avoid confusion we do not introduce any symbol for the dyads of the number eigenvectors),
\be
O = \sum_{m,n=0}^{+\infty} O_{m,n} \vert m \rangle \langle n \vert.
\label{Oa}
\ee
Using the mappings (\ref{bases1}) and (\ref{bases2}), the straightforward calculation shows that
\be
{\cal O}:= BOA= \sum_{m,n=0}^{+\infty} \sqrt{(E_m -\epsilon)(E_n -\epsilon)} \, O_{m,n} X^{m+1,n+1},
\label{Ob}
\ee
and 
\be
\overline {\cal O}:= A^{\dagger} O B^{\dagger}= \sum_{m,n=0}^{+\infty} \sqrt{(E_m -\epsilon)(E_n -\epsilon)} \, O_{m,n}  \overline X^{\, m+1,n+1}.
\label{Oc}
\ee
Thus, in any case $O$ is transformed into the same square matrix. The latter is of infinite order and has its first row and first column with all the entries equal to zero. The action of ${\cal O}$ and $\overline {\cal O}$ on their respective representation spaces is easily derived by using the relationships (\ref{hubbard2}). Notice also that ${\cal O}$ and $\overline {\cal O}$ have the same expectation value $\langle \overline \xi_{\lambda} \vert {\cal O} \vert \xi_{\lambda} \rangle= \langle \xi_{\lambda} \vert \overline {\cal O} \vert \overline \xi_{\lambda} \rangle$. In turn, the adjoint of $O \in \mbox{Aut}({\cal H})$ is transformed into the operators ${\cal O}^+ := B O^{\dagger} A$ and $\overline {\cal O}^{\, +} := A^{\dagger} O^{\dagger} B^{\dagger}$. Therefore:
\be
{\cal O}^{\dagger} = \overline {\cal O}^{\, +}, \quad \overline {\cal O}^{\, \dagger} = {\cal O}^+, \quad \left( {\cal O}^+ \right)^{\dagger}= \overline {\cal O}, \quad \left( \overline {\cal O}^{\, +} \right)^{\dagger}=  {\cal O}.
\ee
That is, ${\cal O}$ and its adjoint ${\cal O}^{\dagger}$ are not elements of the same set since ${\cal O} \in \mbox{Aut} ({\cal H}_{\lambda} )$ and ${\cal O}^{\dagger} \in \mbox{Aut} (\overline {\cal H}_{\lambda} )$. Similarly $\overline {\cal O} \in \mbox{Aut} (\overline {\cal H}_{\lambda} )$ and $\overline {\cal O}^{\dagger} \in \mbox{Aut}  ({\cal H}_{\lambda} )$. These results are consistent with the bi-product (\ref{prod1}), for if we calculate the expectation value of ${\cal O}^{\dagger}$, then we obtain $ \langle {\cal O}^{\dagger} \rangle = \langle \xi_{\lambda} \vert \overline {\cal O}^{\, +} \vert \overline \xi_{\lambda} \rangle=  \langle \overline \xi_{\lambda} \vert  {\cal O} \vert \xi_{\lambda} \rangle^* = \langle {\cal O} \rangle^*$. In particular, if $O$ is self-adjoint then ${\cal O} = {\cal O}^+$, and $ \langle \overline {\cal O} \rangle = \langle {\cal O} \rangle^*$.

\subsection{Basic operators}

We would like to emphasize that the mappings (\ref{map1}) and (\ref{map2}) transform diagonal matrices into diagonal matrices. Then, from the operators in $\mbox{Aut}({\cal H})$ that are represented by diagonal matrices we obtain a first class of operators in $\mbox{Aut}({\cal H}_{\lambda} )$ and $\mbox{Aut}(\overline {\cal H}_{\lambda} )$. For instance, the identity operator $ \mathbb I \in \mbox{Aut}({\cal H})$ gives rise to the same diagonal matrix
\be
{\cal I} \equiv H_{\lambda} -\epsilon = \sum_{n=0}^{+\infty} (E_n -\epsilon) X^{n+1,n+1}, \quad \overline {\cal I} \equiv \overline H_{\lambda} -\epsilon = \sum_{n=0}^{+\infty} (E_n -\epsilon) \overline X^{\, n+1,n+1}.
\ee
A second example is given by the Hamiltonian $H \in \mbox{Aut}({\cal H})$, which also leads to a same diagonal matrix in $\mbox{Aut}({\cal H}_{\lambda} )$ and $\mbox{Aut}(\overline {\cal H}_{\lambda} )$:
\be
\begin{aligned}
{\cal E}:= BHA \equiv H_{\lambda} (H_{\lambda} -\epsilon) = \sum_{n=0}^{+\infty} E_n  (E_n -\epsilon)  X^{n+1,n+1}, 
\\[1ex]
\overline {\cal E}:= A^{\dagger} H B^{\dagger} \equiv \overline H_{\lambda} (\overline H_{\lambda} -\epsilon) = \sum_{n=0}^{+\infty} E_n  (E_n -\epsilon)  \overline X^{ \, n+1,n+1}.
\label{htrans}
\end{aligned}
\ee
Another useful result is obtained from the transformation of the off-diagonal ladder operators $\hat a$ and $\hat a^{\dagger}$. Using $a_{m,n}= \sqrt{2n} \delta_{m,n-1}$ and $a_{m,n}^{\dagger}= \sqrt{2(n+1)} \delta_{m,n+1}$, we have
\be
\begin{aligned}
{\cal A}:= B \hat a A= \sum_{n=0}^{+\infty} g_{\cal A} (n) \, X^{n+1,n+2},\quad 
{\cal A}^+:= B \hat a^{\dagger} A= \sum_{n=0}^{+\infty} g_{\cal A} (n) \, X^{n+2,n+1},\\
\overline {\cal A}:= A^{\dagger} \hat a B^{\dagger}= \sum_{n=0}^{+\infty} g_{\cal A} (n) \,\overline X^{\, n+1,n+2},\quad 
\overline {\cal A}^+:= A^{\dagger}  \hat a^{\dagger}  B^{\dagger}= \sum_{n=0}^{+\infty} g_{\cal A} (n) \, \overline X^{\, n+2,n+1},
\end{aligned}
\label{natural}
\ee
where
\be
g_{\cal A} (n)=\sqrt{2(n+1)(E_n -\epsilon)(E_{n+1}-\epsilon)}.
\label{funcion}
\ee
The action of these operators on the basis elements of their respective representation spaces is as follows
\be
\begin{aligned}
{\cal A} \vert \psi_{n+2} \rangle = g_{\cal A} (n) \vert \psi_{n+1} \rangle, \qquad {\cal A}^+ \vert \psi_{n+1} \rangle = g_{\cal A} (n) \vert \psi_{n+2} \rangle,\\[1ex]
\overline {\cal A} \vert \overline \psi_{n+2} \rangle = g_{\cal A} (n) \vert \overline \psi_{n+1} \rangle, \qquad \overline {\cal A}^{\, +} \vert \overline \psi_{n+1} \rangle = g_{\cal A} (n) \vert \overline \psi_{n+2} \rangle,
\label{action1}
\end{aligned}
\ee
and
\be
{\cal A} \vert \psi_1 \rangle = 
 \overline {\cal A} \vert \overline \psi_1 \rangle =
{\cal A} \vert \psi_0 \rangle = 
 \overline {\cal A} \vert \overline \psi_0 \rangle =  
{\cal A}^+ \vert \psi_0 \rangle = 
 \overline {\cal A}^{\, +} \vert \overline \psi_0 \rangle =
 \vert \varnothing \rangle.
\label{action2}
\ee
According to the above equations, the space of states ${\cal H}_{\lambda}$ is naturally decomposed into the direct sum of two invariant subspaces, the one spanned by $\{ \vert \psi_n \rangle \}_{n\geq 1}$ and that spanned by $\vert \psi_0 \rangle$. In the former case ${\cal A}$ and ${\cal A}^+$ work respectively as annihilation and creation operators, see Fig.~\ref{f_spectra}. In turn, the subspace spanned by $\vert \psi_0 \rangle$ is in the intersection of the kernels of ${\cal A}$ and ${\cal A}^+$ since such vector is annihilated by both operators. A similar description can be done for $\overline {\cal H}_{\lambda}$ with respect to the operators $\overline {\cal A}$ and $\overline {\cal A}^{\, +}$.

From (\ref{natural}), the transformation of the position and momentum operators is easily obtained 
\be
\begin{aligned}
X:= B \hat x A = \tfrac12 \left( {\cal A}^+ + {\cal A} \right), \qquad P:= B \hat p A = \tfrac{i}2 \left( {\cal A}^+ - {\cal A} \right),\\[1ex]
\overline X:= A^{\dagger} \hat x B^{\dagger}  = \tfrac12 \left( \overline {\cal A}^+ +\overline {\cal A} \right), \qquad \overline P:= A^{\dagger}  \hat p B^{\dagger}  = \tfrac{i}2 \left( \overline {\cal A}^+ - \overline {\cal A} \right).
\end{aligned}
\label{XP}
\ee
Their action on the corresponding representation spaces can be calculated from (\ref{action1}) and (\ref{action2}). The quadratic forms of the above operators lead to
\be
P^2 + X^2= B[H(H-\epsilon)+2]A=[H_{\lambda}(H_{\lambda}-\epsilon) +2](H_{\lambda}-\epsilon),
\label{quad1}
\ee
and a similar expression for the operators in $\mbox{Aut}( \overline {\cal H}_{\lambda})$. Remark that $P^2 + X^2$ corresponds to the transformation of the quadratic polynomial $H(H-\epsilon) +2$. Therefore, although its matrix-representation is diagonal, this operator does not correspond to the Hamiltonian $H_{\lambda}$ but to the cubic polynomial of $H_{\lambda}$ appearing in Eq.~(\ref{quad1}). 

\section{Algebras of operators}
\label{algebras}

The Hubbard operators (\ref{hubbard}) facilitate the calculation of the commutators in $\mbox{Aut} ({\cal H}_{\lambda})$ and $\mbox{Aut} (\overline {\cal H}_{\lambda})$. They are used in the next sections to identify the algebras that are obeyed by the ladder operators associated with the non-Hermitian oscillators $H_{\lambda}$ and $\overline H_{\lambda}$.

\subsection{Quadratic polynomial Heisenberg algebras}
\label{quadratic}

The commutator between ${\cal A}$ and ${\cal A}^+$ yields the diagonal matrix
\be
[{\cal A}, {\cal A}^+] = 
\sum_{n=0}^{+\infty} 2 (3E_n -\epsilon) (E_n -\epsilon)  X^{n+1,n+1}.
\ee
Comparing with (\ref{htrans}) we realize that $[{\cal A}, {\cal A}^+]$ must correspond to the transformation of a function of the Hamiltonian $H$. After some simple operations we find
\be
[{\cal A}, {\cal A}^+] \equiv B 2 (3H-\epsilon)A= 2 (3H_{\lambda}-\epsilon) (H_{\lambda}-\epsilon).
\label{Acomm1}
\ee
That is, the commutator $[{\cal A}, {\cal A}^+]$  is equal to the transformation of the operator $2(3 H-\epsilon)$. The latter is the reason for which (\ref{Acomm1}) is a quadratic polynomial of $H_{\lambda}$. The following commutation rules can be also verified
\be
[H_{\lambda}, {\cal A} ]=-2 {\cal A}, \qquad [H_{\lambda}, {\cal A}^+ ]=2 {\cal A}^+.
\label{Acomm2}
\ee
Therefore, ${\cal A}$, ${\cal A}^+$ and $H_{\lambda}$ are the generators of a quadratic polynomial algebra defined by the rules (\ref{Acomm1})-(\ref{Acomm2}). In similar form, $\overline {\cal A}$, $\overline {\cal A}^{\, +}$ and $\overline H_{\lambda}$ are the generators of the quadratic polynomial algebra
\be
[\overline {\cal A}, \overline {\cal A}^{\, +}] = 2 (3 \overline H_{\lambda}-\epsilon) (\overline H_{\lambda}-\epsilon), \quad
[\overline H_{\lambda}, \overline {\cal A} ]=-2 \overline {\cal A}, \quad [\overline H_{\lambda}, \overline {\cal A}^{\, +} ]=2 \overline {\cal A}^{\, +}.
\label{Acomm3}
\ee
The polynomial algebras are quite natural in the higher order supersymmetric approaches \cite{Mie04,Aoy01a,Aoy01b,And12}. They are usually connected with nonlinearities that arise because the differential order of the operators that intertwine the susy partner Hamiltonians is greater than one. In the present case, the algebra is quadratic polynomial because ${\cal A}$ and ${\cal A}^{\dagger}$ are the result of transforming the first-order differential operators $\hat a$ and $\hat a^{\dagger}$ by the action of two additional first-order differential operators. Indeed, the algebra (\ref{Acomm1})-(\ref{Acomm2}) is also associated with the Hermitian oscillator-like Hamiltonians reported in \cite{Mie84}. The latter means that the bi-orthogonal system formed by the eigenvectors of the non-Hermitian oscillators presented here and the space of states introduced in \cite{Mie84} are two different representation spaces of the same polynomial algebra.

Now, using (\ref{XP}) we also obtain the commutation rules for the transformed position and momentum operators
\be
[ X, P] = i (3H_{\lambda}-\epsilon) (H_{\lambda}-\epsilon), \quad [H_{\lambda},  X]= -2i  P, \quad [H_{\lambda},  P]= 2i  X,
\label{Acomm4}
\ee
with similar expressions for the operators in $\mbox{Aut}( \overline {\cal H}_{\lambda})$. The algebra (\ref{Acomm4}) is quadratic in $H_{\lambda}$, as it would be expected because it is associated with (\ref{Acomm1})-(\ref{Acomm2}). In turn, the root-mean-square deviations $\Delta X =\sqrt{ \langle X^2 \rangle -  \langle X \rangle^2}$ and $\Delta P =\sqrt{ \langle P^2 \rangle -  \langle P \rangle^2}$ satisfy the inequality
\be
\Delta X \Delta P \geq \tfrac12 \vert \langle \left( 3H_{\lambda} -\epsilon \right) \left( H_{\lambda} -\epsilon \right) \rangle \vert.
\label{ineq1}
\ee

Notice that the {\em polynomial Heisenberg algebras} introduced above depend on the ground energy of the system. That is, systems with different ground energies $\epsilon$ will be regulated by different algebras. The latter seems to be quite natural if one considers that it is the energy spectrum which characterizes a one-dimensional system in general. For instance, the systems represented by the potentials shown in Figs.~\ref{f_pot}(a) and \ref{f_pot}(d) obey the same polynomial algebra, but this last is different from the algebra associated with the potentials depicted in either Figs.~\ref{f_pot}(b) and \ref{f_pot}(e),  or Figs.~\ref{f_pot}(c) and \ref{f_pot}(f). However, the feature that defines the families of operators $H_{\lambda}$ and $\overline H_{\lambda}$ is not the ground energy $\epsilon$ by itself, but the fact that the corresponding eigenvectors, $\vert \psi_0 \rangle$ and $\vert \overline \psi_0 \rangle$, are annihilated by ${\cal A}$ as well as by ${\cal A}^+$. In other words, no matter the value of $\epsilon$, no eigenvector $\vert \psi_{n+1} \rangle$ is connected with $\vert \psi_0 \rangle$ via the ladder operators. The same is true for $\vert \overline \psi_{n+1} \rangle$ and $\vert \overline \psi_0 \rangle$. As discussed above, this property is a natural consequence of the form in which the mappings (\ref{map1}) and (\ref{map2}) have been designed. We wonder about the possibility of finding a simpler algebra for such a striking profile. In the next section we discuss in detail about the point.

\subsection{Distorted Heisenberg algebras} 
\label{distorted}

Let us look for ladder operators satisfying simple algebras in $\mbox{Aut}( {\cal H}_{\lambda})$ and $\mbox{Aut}( \overline {\cal H}_{\lambda})$. Consider the $f$-boson operators \cite{Man97}
\be
\hat a_f = \hat a f(\hat N) = f(\hat N+1) \hat a, \quad 
\hat a_f^{\dagger} = f(\hat N) \hat a^{\dagger} = a^{\dagger} f(\hat N+1),
\label{f-osc}
\ee
where $f$ is a real (self-adjoint) function of the number operator $\hat N$. The above operators satisfy the commutation relation $[\hat a_f, \hat a_f^{\dagger}]=2 (\hat N +1)f^2 (\hat N+1) -2\hat N f^2(\hat N)$, and are transformed as follows
\be
{\cal C}= B \hat a_f A, \quad {\cal C}^+= B \hat a_f ^{\dagger} A,
\label{dist1}
\ee
with similar expressions for the operators in $\mbox{Aut}( \overline {\cal H}_{\lambda})$. The commutator between ${\cal C}$ and ${\cal C}^+$ yields $[ {\cal C}, {\cal C}^+] = B F(H) A$, where
\be
F(H)  =  (H+1)  f^2 (\hat N+1)  (H+2-\epsilon) - (H-1)f^2 (\hat N) (H-2-\epsilon).
\label{rop}
\ee
In matrix-representation the above operator acquires the diagonal form
\be
[ {\cal C}, {\cal C}^+] = \sum_{n=0}^{+\infty} F(E_n) (E_n -\epsilon) X^{n+1,n+1}.
\label{rcom}
\ee
With the exception of the first term in the sum, the coefficients of (\ref{rcom}) have the structure
\be
F(E_{n+1}) (E_{n+1} -\epsilon) = R(n+1) -R(n), \quad n \geq 0,
\label{star}
\ee
where
\be
R(n) = 2(n+1) f^2(n+1) (2n+3-\epsilon) (2n+1-\epsilon), \quad n \geq 0.
\ee
In turn, assuming that $f(0)$ is finite,  the coefficient of the first term in (\ref{rcom}) is as follows
\be
F(E_0)(E_0-\epsilon)= R(0).
\label{rcero}
\ee
To determine $f (\hat N)$ let us impose the condition
\be
F(E_{n+1}) (E_{n+1} -\epsilon) =\kappa =\mbox{const} \quad \forall \, n \geq 0.
\label{finite1}
\ee
From (\ref{star}) we see that (\ref{finite1}) is actually a difference equation. To simplify the expressions we use $R(n) = \kappa S(n)$. Then
\be
S(n+1)-S(n)= 1, \quad n \geq 0.
\label{finite2}
\ee
Proposing $S(n)=w \alpha^n$ as a solution of the homogeneous equation $S(n+1) -S(n)=0$ we obtain $w\alpha^n (\alpha -1)=0$ for all non-negative integer $n$, so that (\ref{finite2}) is a difference equation of first order and $\alpha=1$. The latter means that the simplest solution of (\ref{finite2}) is a sequence of consecutive non-negative integers $S(n)=n$. Therefore, the general solution is of the form $S(n)= w + n$. To fix the initial condition we take $S(0)=w \geq 0$. Therefore,
\be
R(n)=2 (n+1)  f^2(n+1)(2n+3-\epsilon)(2n+1-\epsilon) = \kappa(w+n), \qquad w, n \geq 0,
\label{r}
\ee
leads to the $w$-parameterized function
\be
f_w (n+1 )=\left[ \frac{ \kappa(w+n) }{ 2 (n+1) (2n +3 -\epsilon) (2n+1-\epsilon)}
\right]^{1/2}, \quad w,n \geq 0.
\label{efe}
\ee
For simplicity, hereafter we take $\kappa=2$. Notice that $f_w (n+1 )$ is a real-valued function of the integer $n+1$, so that the lowest value of its argument is $1$. That is, we do not know the explicit form of $f_w(0)$, although we have assumed that it is finite. Coming back to the Eq.~(\ref{rcero}), now we can write
\be
F(E_0)(E_0-\epsilon)= 2w.
\label{rcero2}
\ee
Using (\ref{finite1}) and (\ref{rcero2}), the commutator (\ref{rcom}) is expressed as
\be
[ {\cal C}_w, {\cal C}_w^+] = 2 I_w,
\label{rcom2}
\ee
where
\be
I_w= w X^{1,1} +  \sum_{n=1}^{+\infty} X^{n+1,n+1}.
\label{Iw}
\ee
On the other hand, after substituting $n$ by $\hat N$ in (\ref{efe}) one arrives at the operator-function
\be
 f_w (\hat N +1)= \left[ \frac{ w + \hat N }{ (\hat N +1)  ( 2 \hat N + 3-\epsilon) (2 \hat N +1 -\epsilon) } \right]^{1/2}.
 \label{efew}
\ee
Then, we have the $f_w$-boson operators 
\be
\hat a_{f_w} = f_w (\hat N+1) \hat a , \quad 
\hat a_{f_w}^{\dagger} =  a^{\dagger} f_w (\hat N+1),
\label{f-wosc}
\ee
which act on the number eigenvectors $\vert n \rangle$ as follows
\be
\hat a_{f_w}^{\dagger} \vert n \rangle = \vartheta_w(n,\epsilon) \vert n+1 \rangle, \quad \hat a_{f_w} \vert n+1 \rangle = \vartheta_w(n,\epsilon) \vert n \rangle, \quad \hat a_{f_w} \vert 0 \rangle = \vert \varnothing \rangle, \quad n\geq 0,
\label{number}
\ee
with
\be
\vartheta_w(n,\epsilon) = \left[ \frac{ 2(w+n)}{ (2n+3-\epsilon)(2n+1-\epsilon) }
\right]^{1/2}.
\label{wteta}
\ee
The $f_w$-boson operators are transformed as
\be
{\cal C}_w = B f_w (\hat N +1) \hat a A, \quad {\cal C}_w^+ = B \, \hat a^{\dagger} f_w (\hat N +1) A.
\label{Cw}
\ee
The action of the transformed operators on the eigenvectors of $H_{\lambda}$ is very simple
\be
{\cal C}_w \vert \psi_{n+2} \rangle = \sqrt{2(w+n)} \,  \vert \psi_{n+1} \rangle, \quad  {\cal C}_w^+ \vert \psi_{n+1} \rangle = \sqrt{2(w+n)} \, \vert \psi_{n+2} \rangle,
\ee
with ${\cal C}_w \vert \psi_1 \rangle = {\cal C}_w \vert \psi_0 \rangle = {\cal C}_w^+ \vert \psi_0 \rangle =  \vert \varnothing \rangle$. The straightforward calculation shows that 
\be
[H_{\lambda}, {\cal C}_w ]=-2 {\cal C}_w, \qquad [H_{\lambda}, {\cal C}_w^+ ]=2 {\cal C}_w^+.
\label{dist2}
\ee
From (\ref{rcom2}) and (\ref{dist2}) we see that ${\cal C}_w$, ${\cal C}_w^{\dagger}$ and $I_w$ are the generators of an algebra which imitates the Heisenberg one. For if we concentrate on the subspace spanned by $\{ \vert \psi_n \rangle \}_{n \geq 2}$ only, then the commutators (\ref{rcom2}) and (\ref{dist2}) are completely equivalent to the oscillator ones. The same occurs in the subspace spanned by $\vert \psi_1 \rangle$, up to the constant $w \geq 0$. For this reason, the commutation rules (\ref{rcom2}) and (\ref{dist2}) will be referred to as  {\em distorted Heisenberg algebra}, and $w$ will be called the {\em distortion parameter}. As before, the subspace spanned by $\vert \psi_0 \rangle$ is in the intersection of the kernels of ${\cal C}_w$ and ${\cal C}_w^+$. 

It is worth noting that the distorted Heisenberg algebra was introduced in \cite{Fer95,Ros96} for the Hermitian oscillator-like Hamiltonians reported in \cite{Mie84} (further improvements can be found in \cite{Fer07}). That is, we have two different representation spaces for such algebra, the one given by the bi-orthogonal system presented here and that reported in \cite{Fer95,Ros96}. On the other hand, using the ladder operators (\ref{Cw}) we obtain the {\em distorted quadratures} $X_w = \tfrac12 \left( {\cal C}_w^+ +{\cal C}_w \right)$ and $P_w = \tfrac{i}2 \left( {\cal C}_w^+ - {\cal C}_w \right)$. They satisfy the commutation relationship $[X_w, P_w]= \tfrac{i}2 [{\cal C}_w, {\cal C}_w^+]= i I_w$, and the inequality $\Delta X_w \Delta P_w \geq \tfrac12 \vert \langle I_w \rangle \vert$. The construction of the distorted ladder operators $\overline {\cal C}_w, \overline {\cal C}_w^+ \in \mbox{Aut} ( \overline {\cal H}_{\lambda})$ and the distorted identity $\overline I_w \in \mbox{Aut} ( \overline {\cal H}_{\lambda})$ is achieved in similar form, with the same conclusions as those obtained above.

Let us remark that, contrary to what happens with the algebra of the previous section, the distorted Heisenberg algebra does not depend on the ground energy of the system. In fact, the distorted algebra is a common property of all the non-Hermitian oscillators $H_{\lambda}$ and $\overline H_{\lambda}$. For instance, the spectral properties of all the potentials depicted in Fig.~\ref{f_pot} can be described by using the same distorted algebra, no matter the value of the ground energy $\epsilon$.

\section{Generalized coherent states}
\label{gcs}

In Sec.~\ref{transfT} we have shown that the mappings (\ref{map1}) and (\ref{map2}) define the transformation of an arbitrary vector $\vert \phi \rangle \in {\cal H}$, see Eq.~({\ref{tlin1}), into either $\vert \phi^{(\lambda)} \rangle \in {\cal H}_{\lambda}$ or $\vert \overline \phi^{(\lambda)} \rangle \in \overline{\cal H}_{\lambda}$, Eqs.~(\ref{tlin2}) and (\ref{tlin3}) respectively. To show the relevance of such transformations let us make $\vert \phi \rangle$ equal to the coherent state of the mathematical oscillator:
\[
\vert \alpha \rangle = e^{-\frac{\vert \alpha \vert^{2}}{4}} \sum_{n=0}^{+\infty} \frac{\left( \alpha/\sqrt{2} \right)^{n}}{\sqrt{n!}} \, \vert n \rangle.
\]
That is, using $\vert \phi \rangle = \vert \alpha \rangle$ in (\ref{tlin2}) we obtain
\be
\vert \alpha^{(\lambda)} \rangle = \frac{ e^{\frac{-\vert \alpha \vert^2}{4} } }{ \sqrt{ \vert \alpha \vert^2 +1 -\epsilon} } \sum_{n=0}^{\infty} \sqrt{ \frac{ 2n+1-\epsilon}{n!} } \left( \frac{\alpha}{\sqrt 2} \right)^ n \vert \psi_{n+1} \rangle.
\label{pad1}
\ee
The expression (\ref{tlin3}) for the concomitant $\vert \overline \alpha^{(\lambda)} \rangle$ is obtained from the above equation by changing $\vert \psi_{n+1} \rangle$ to $\vert \overline\psi_{n+1} \rangle$. The structure of the transformed coherent state (\ref{pad1}) resembles that of the one-photon added coherent state for the oscillator \cite{Aga91,Siv99}. Indeed, for $\epsilon=-1$ one has
\be
\left. \vert \alpha^{(\lambda)} \rangle \right\vert_{\epsilon =-1} = \frac{ e^{\frac{-\vert \alpha \vert^2}{4} } }{ \sqrt{ \vert \alpha \vert^2 + 2} } \sum_{n=0}^{\infty} \sqrt{ \frac{ 2(n+1)}{n!} } \left( \frac{\alpha}{\sqrt 2} \right)^ n \vert \psi_{n+1} \rangle,
\label{pad2}
\ee
which coincides with the `mathematical' form of the states reported in \cite{Aga91,Siv99} after changing $\vert \psi_n \rangle$ to $\vert n \rangle$. That is, for $\epsilon =-1$ the coherent state $\vert \alpha \rangle \in {\cal H}$ is mapped into ${\cal H}_{\lambda}$ as the `one-photon added coherent state' (\ref{pad2}), see Sec.~\ref{equidistant} for details. 

In addition to the transformed coherent states (\ref{pad1}), one can use the algebras introduced in the previous section to construct different sets of well defined coherent states.

\subsection{Natural coherent states}
\label{natCS}

Consider the quadratic polynomial Heisenberg algebra derived in Sec.~\ref{quadratic}. We want to solve the eigenvalue equation ${\cal A} \vert \phi^{({\cal N})} \rangle = z \vert  \phi^{({\cal N})} \rangle$, with $z$ a complex number. From (\ref{action2}) we know that the eigenvalue $z=0$ is twice degenerate since $\vert \psi_0 \rangle$ and $\vert \psi_1 \rangle$ are annihilated by ${\cal A}$. For $z\neq 0$, the vector we are looking for may be written as a normalized bi-orthogonal superposition of the eigenvectors $\vert \psi_n \rangle$. Indeed, from (\ref{action1}) we obtain
\be
\vert  \phi^{({\cal N})} (z) \rangle = c^{({\cal N})}_0 (\vert z \vert)
\sum_{n=0}^{+\infty} c_{n+1}^{({\cal N})}  (z) \vert \psi_{n+1} \rangle, \quad z\in \mathbb C,
\label{ec1}
\ee
with 
\be
c_{n+1}^{({\cal N})} (z)  = \left[ \frac{\Gamma \left( \frac{1- \epsilon}{2} \right)  \Gamma \left(\frac{3- \epsilon}{2} \right) }{ n! \, \Gamma \left( n + \frac{1-\epsilon}{2} \right) \Gamma \left( n + \frac{3-\epsilon}{2} \right) } \right]^{1/2}  \left( \frac{z}{\sqrt{8} } \right)^n
, \quad z\in \mathbb C,
\label{ec2}
\ee
and
\be
c^{({\cal N})}_0 (\vert z \vert) =\left[ {}_0F_2 \left( \frac{1-\epsilon}{2}, \frac{3-\epsilon}{2}; \frac{\vert z \vert^2}{8} \right) \right]^{-1/2}
\label{ec3}
\ee 
the normalization constant. Clearly, $\vert  \phi^{({\cal N})} (z=0) \rangle= \vert \psi_1 \rangle$. Then, the set $\{ \vert \psi_0 \rangle, \vert \phi^{({\cal N})} (z) \rangle \}$ includes all the normalized eigenvectors of ${\cal A}$ with complex eigenvalue $z$. In turn, it is straightforward to show that $\{ \vert \overline \psi_0 \rangle, \vert \overline \phi^{({\cal N})} (y) \rangle \}$ is the set of all the normalized eigenvectors of $\overline {\cal A}$ with complex eigenvalue $y$. Moreover, $\vert \overline \phi^{({\cal N})} (y) \rangle$ and $\vert \phi^{({\cal N})} (z) \rangle$ are concomitant if $y=z$. For if we calculate their product
\be
K_{\cal N}(y^*,z):= \langle \overline \phi^{({\cal N})} (y) \vert \phi^{({\cal N})} (z) \rangle 
=  \frac{ {}_0F_2 (  \frac{1-\epsilon}{2}, \frac{3-\epsilon}{2}; zy^*/8)
}{
\sqrt{ {}_0F_2 ( \frac{1-\epsilon}{2}, \frac{3-\epsilon}{2}; \vert y \vert^2/8) \, {}_0F_2 ( \frac{1-\epsilon}{2}, \frac{3-\epsilon}{2}; \vert z \vert^2/8) 
}},
\label{repro1}
\ee
then we see that $y=z$ implies $K_{\cal N}(z^*,z)=1$. On the other hand, using (\ref{ortho1}) one can verify that
\be
\langle \overline \psi_0 \vert \phi^{({\cal N})} (z) \rangle = \langle \overline \phi^{({\cal N})} (z) \vert \psi_0 \rangle =0.
\ee
Therefore we may write the closure relation $\mathbb I_{\lambda}$ as
\be
\mathbb I_{\lambda}= \vert \psi_0 \rangle \langle \overline \psi_0 \vert + \int \vert \phi^{({\cal N})} (z) \rangle \langle \overline \phi^{({\cal N})} (z) \vert \, d\mu^{({\cal N})} (z),
\label{int1}
\ee
where the measure $d\mu^{({\cal N})} (z)$ is to be determined. Assuming dependence on $\vert z \vert$ only, we can write
\be
d \mu^{({\cal N})} (z) = \frac{ {}_0F_2 \left( \frac{1-\epsilon}{2}, \frac{3-\epsilon}{2};  \frac{r^2}{8} \right) }
{\Gamma \left( \frac{1-\epsilon}{2} \right) \Gamma \left( \frac{3-\epsilon}{2} \right) } \, h(r^2) rdr d \phi, \quad z= re^{i\phi},
\label{mu}
\ee
with $h(r^2)$ a new function to be determined. After integrating over the angular variable, the introduction of (\ref{mu}) into (\ref{int1}) yields
\be
\mathbb I_{\lambda}= \vert \psi_0 \rangle \langle \psi_0 \vert + \sum_{n=0}^{+\infty} \Lambda_n \vert \psi_{n+1} \rangle \langle \overline \psi_{n+1} \vert,
\label{int2}
\ee
where
\be
\Lambda_n=\frac{ 8\pi 
}{ n! \, \Gamma \left( n + \frac{1-\epsilon}{2} \right) \Gamma \left( n + \frac{3-\epsilon}{2} \right) } \int_0^{+\infty} x^n h(x) dx, \quad n\geq 0.
\label{int3}
\ee
To obtain $h(x)$ let us impose the condition $\Lambda_n=1$, which is sufficient to make (\ref{int2}) equal to (\ref{iden3}). Hence, after the change $n \rightarrow m-1$, we arrive at the integral equation
\be
 \int_0^{+\infty} x^{m-1} [8\pi  h(x)] dx = 
 \Gamma(m)  \Gamma \left( m - \tfrac{1+\epsilon}{2} \right) \Gamma \left( m+ \tfrac{1-\epsilon}{2} \right), \quad m \geq 1,
 \label{int4}
\ee
which is the Mellin transform of $h(x)$ \cite{Ber00}. Thus, the measure (\ref{mu}) is determined by the Mellin-Barnes integral representation \cite{Olv10} of the following Meijer G-function 
\be
8 \pi h(x) = G_{0,3}^{3,0} \left(
\begin{array}{c|c}
x & \begin{array}{c}
 -\\[1ex]
0, -\frac{1+\epsilon}{2}, \frac{1-\epsilon}{2} 
\end{array}
\end{array}
\right).
\ee
Provided $h(x)$, the identity (\ref{int1}) permits the decomposition of $\vert \phi^{({\cal N})} (z) \rangle$ as follows 
\be
\vert \phi^{({\cal N})} (z) \rangle = \int K_{\cal N}(y^*,z)  \vert \phi^{({\cal N})} (y) \rangle \, d\mu^{({\cal N})} (y) ,
\ee
so that the function $K_{\cal N}(y^*,z)$ introduced in (\ref{repro1}) is the reproducing kernel
\be
K_{\cal N} (\eta^*,z) =  \int K_{\cal N} (y^*,z) K_{\cal N} (\eta^*,y) \, d\mu^{({\cal N})} (y).
\ee
Besides, given any $\vert \xi_{\lambda} \rangle \in {\cal H}_{\lambda}$ we have
\be
\vert \xi_{\lambda} \rangle = c_0 \vert \psi_0 \rangle+ \int  \xi_{\lambda}(z, z^*) \vert \phi^{({\cal N})} (z) \rangle \, d \mu^{({\cal N})} (z) 
\ee
where
\be
\xi_{\lambda} (z,z^*):= \langle \overline \phi^{({\cal N})} (z) \vert \xi_{\lambda} \rangle = c_0^{*({\cal N})} (\vert z \vert) 
 \sum_{n=0}^{+\infty} c_{n+1}^{*({\cal N})} (z) \, c_{n+1}
\ee
is a complex-valued function determined by the coefficients (\ref{ec2}) and (\ref{ec3}).

The expectation value of the Hamiltonian $H_{\lambda}$ in terms of the vectors (\ref{ec1})-(\ref{ec3}) is as follows
\be
\langle H_{\lambda} \rangle_z = 1 + \frac{ r^2}{(1-\epsilon) (3-\epsilon)} \frac{ {}_0F_2 \left( \frac{3-\epsilon}{2}, \frac{5-\epsilon}{2};  \frac{r^2}{8} \right)
}
{{}_0F_2 \left( \frac{1-\epsilon}{2}, \frac{3-\epsilon}{2};  \frac{r^2}{8} \right)},
\ee
and the variances of $X$ and $P$ are given by
\be
\Delta X^2= \Delta P^2 \equiv \Delta X \Delta P = \tfrac12  \langle [X, P] \rangle_z = \tfrac12 \left[ 3 \langle H_{\lambda}^2 \rangle_z - 4 \epsilon \langle H_{\lambda} \rangle_z + \epsilon^2 \right]
\label{variances}
\ee
with 
\be
\langle H_{\lambda}^2  \rangle_z = 4 \langle H_{\lambda} \rangle_z
-3  + \frac{r^4 }{ (1-\epsilon) (3-\epsilon)^2 (5-\epsilon) } \frac{ {}_0F_2 \left( \frac{5-\epsilon}{2}, \frac{7-\epsilon}{2};  \frac{r^2}{8} \right)
}
{{}_0F_2 \left( \frac{1-\epsilon}{2}, \frac{3-\epsilon}{2};  \frac{r^2}{8} \right)}.
\ee
That is, the inequality (\ref{ineq1}) is minimized by the vectors $\vert \phi^{({\cal N})} (z) \rangle$. In Fig.~\ref{f_natural} we show the behavior of the variances (\ref{variances}) associated with the cases depicted in Fig.~\ref{f_pot}. Notice that the lowest value is reached at $z=0$ because $\vert \phi^{({\cal N})} (z=0) \rangle = \vert \psi_1 \rangle$ in all the cases. In general, no matter the value of $\epsilon$, the variances increase their value as $r \rightarrow \infty$ . On the other hand, the larger the value of $\vert \epsilon \vert$, the greater the uncertainty $\Delta X \Delta P$.

\begin{figure}[htb]
\centering

\centering
{\includegraphics[width=0.3\textwidth]{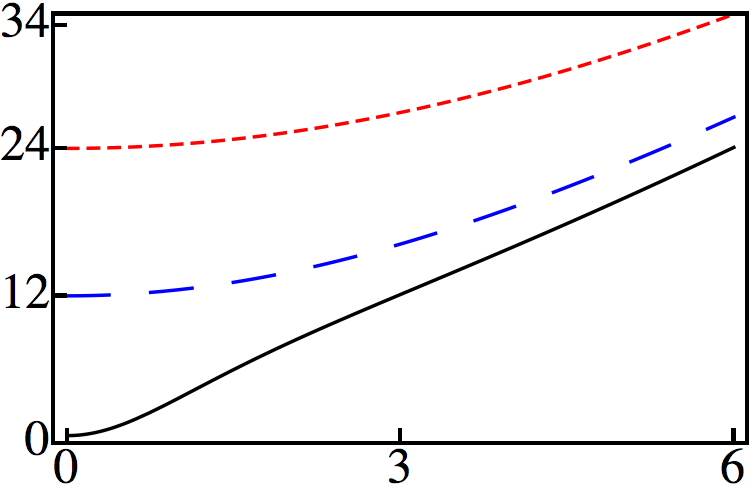}} 

\caption{\footnotesize 
(Color online) The variances $\Delta X^2= \Delta P^2 \equiv \Delta X \Delta P$ defined in (\ref{variances}) for $\epsilon = 0.5$ (solid, black), $\epsilon =-3$ (dashed, blue), and $\epsilon=-5$ (short dashed, red) in terms of $r=\vert z \vert$. In all cases, at $r=0$ the uncertainty takes the value $\Delta X \Delta P = \tfrac12 (3-4\epsilon+\epsilon^2)$. See complementary information in Fig.~\ref{f_pot}.
}
\label{f_natural}
\end{figure}

In summary, all the elements in the set $\{ \vert \psi_0 \rangle, \vert \phi^{({\cal N})} (z) \rangle \}$ are generalized coherent states for the non-Hermitian oscillator $H_{\lambda}$. We shall call them {\em natural coherent states}. Similar conclusions are valid for the set $\{ \vert \overline \psi_0 \rangle, \vert \overline \phi^{({\cal N})} (z) \rangle \} \subset \overline {\cal H}_{\lambda}$.

\subsection{Distorted coherent states}
\label{distCS}

From the algebra introduced in Sec.~\ref{distorted} one finds that the solutions of the eigenvalue equation ${\cal C}_w \vert \phi^{(w)} (z) \rangle = z \vert \phi^{(w)} (z)  \rangle$ are the vectors $\vert \psi_0 \rangle$ and $\vert \psi_1 \rangle$, both of them with eigenvalue $z=0$, and the normalized vectors defined by the bi-orthogonal superposition
\be
\vert \phi^{(w)} (z)  \rangle = c_0^{(w)} (\vert z \vert)  \sum_{n=0}^{+\infty} c_{n+1}^{(w)} (z) \vert \psi_{n+1} \rangle,
\label{distcoh}
\ee
with 
\be
c_{n+1}^{ (w) } (z) = \left[ \frac{\Gamma(w) }{ \Gamma(w+n)}
\right]^{1/2} \left( \frac{z}{\sqrt{2}} \right)^n,
\label{cdist}
\ee
and
\be
c_0^{(w)} (\vert z \vert)  = \left[ {}_1F_1 \left( 1,w; \frac{\vert z \vert^2}{2} \right) \right]^{-1/2}
\label{constdist}
\ee
the normalization constant. The straightforward calculation shows that the new set $\{ \vert \psi_0 \rangle, \vert \phi^{(w)} (z)  \rangle \}$ satisfies the closure relation
\be
\mathbb I_{\lambda} = \vert \psi_0 \rangle \langle \overline \psi_0 \vert + \int  \vert \phi^{(w)} (z)  \rangle \langle \overline\phi^{\, (w)} (z)  \vert \, d\mu^{(w)} (z),
\ee
where $z= re^{i\theta}$ and
\be
d\mu^{(w)} (z) = \frac{r^{2 (w-1)} }{\pi \Gamma (w) 2^w} e^{-r^2/2} {}_1 F_1 (1,w; r^2/2) r dr d\theta
\ee
is the measure, which has been derived by following the same procedure as in the previous case. The reproducing kernel is in this case
\be
K_w(y^*,z):= \langle \overline \phi^{(w)} (y)  \vert \phi^{(w)} (z)  \rangle  = \frac{ {}_1F_1 (1,w; y^* z /2) }{ \sqrt{ {}_1F_1 (1,w; \vert y \vert^2/2)  {}_1F_1 (1,w; \vert z \vert^2/2) }
}
\ee
and the variances of $X_w$ and $P_w$ are of the form
\be
\begin{aligned}
\Delta X_w^2  = \Delta P_w^2 & \equiv \Delta X_w \Delta P_w = \tfrac12  \langle I_w \rangle_z \\[2ex]
& = \frac12 \left[
w - r^2 +\frac{r^2}{w} \frac{ {}_1F_1 (2,w+1; r^2/2)
}{ {}_1F_1 (1,w; r^2/2) }
\right].
\end{aligned}
\label{distvaria}
\ee
Therefore, the vectors (\ref{distcoh}) represent minimum uncertainty states. However, in contrast with the results of the previous case (see Fig.~\ref{f_natural}), the behavior of the distorted variances (\ref{distvaria}) does not depend on the ground energy $\epsilon$ of the system. In fact, given any $\epsilon < E_0 =1$, the uncertainty $\Delta X_w \Delta P_w$ can be manipulated by tuning the distortion parameter $w \geq 0$, as it is shown in Fig.~\ref{f_distorted}.

\begin{figure}[htb]
\centering

\centering
{\includegraphics[width=0.3\textwidth]{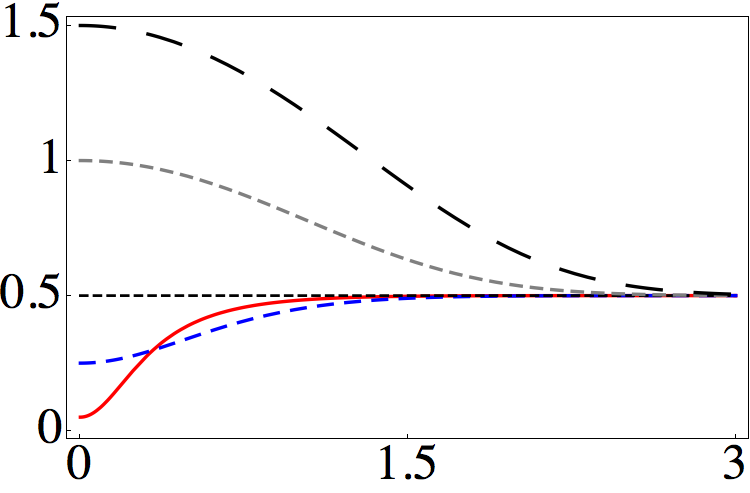}} 

\caption{\footnotesize 
(Color online) The variances $\Delta X_w^2= \Delta P^2 \equiv \Delta X_w \Delta P_w$ defined in (\ref{distvaria}) for $w = 0.1$ (solid, red), $w = 0.5$ (dashed, blue), $w =1$ (horizontal, dotted), $w=2$ (short dashed, gray), and $w=3$ (large dashed, black) in terms of $r= \vert z \vert$. In all cases, at $r=0$ the uncertainty takes the value $\Delta X_w \Delta P_w = \tfrac{w}2$. Compare with Fig.~\ref{f_natural}.
}
\label{f_distorted}
\end{figure}

In summary, all the elements in the set $\{ \vert \psi_0 \rangle, \vert \phi^{(w)} (z) \rangle \}$ are generalized coherent states for the non-Hermitian oscillator $H_{\lambda}$. We shall call them {\em distorted coherent states}. Similar conclusions are valid for the set $\{ \vert \overline \psi_0 \rangle, \vert \overline \phi^{(w)} (z) \rangle \} \subset \overline {\cal H}_{\lambda}$.

\section{Continuous representation}
\label{continuo}

Let us introduce a unified notation for the families of generalized coherent states derived in the previous section
\be
\vert \phi^{(s)} (z) \rangle= c_0^{(s)} (\vert z \vert) \vert \phi^{(s)} (z) \rangle_U, \quad s={\cal N}, w,
\label{csgral}
\ee
where
\be
\vert \phi^{(s)} (z) \rangle_U= \sum_{n=0}^{\infty} c_{n+1}^{(s)} (z) \vert \psi_{n+1} \rangle 
\label{csgral2}
\ee
is the unnormalized version of (\ref{csgral}) and the super-index ``$s$'' stands for either natural $({\cal N})$ or distorted $(w)$. The identity operator $\mathbb I_{\lambda} \subset \mbox{Aut}({\cal H}_{\lambda})$ acquires the generic form
\be
\mathbb I_{\lambda}= \vert \psi_0 \rangle \langle \overline\psi_0 \vert + \int \vert \phi^{(s)} (z) \rangle \langle \overline \phi^{(s)} (z) \vert \, d\mu^{(s)} (z),
\label{idengral}
\ee
so that any vector $\vert f_{\lambda} \rangle \in {\cal H}_{\lambda}$ can be written as 
\be
\vert f_{\lambda} \rangle = f_0 \vert \psi_0 \rangle + \int f^{(s)}_{\lambda} (z,z^*) \vert \phi^{(s)}  (z) \rangle \, d\mu^{(s)} (z)
\ee
where $f_0 \equiv \langle \overline \psi_0 \vert f_{\lambda} \rangle$ is a constant and
\be
f^{(s)}_{\lambda} (z, z^*) \equiv \langle \overline \phi^{(s)} (z) \vert f_{\lambda} \rangle = c_0^{ * (s)} (\vert z \vert) \, f_s(z^*),
\label{opt1}
\ee
with
\be
f_s(z^*):=  {}_U\langle \overline \phi^{(s)} (z) \vert f_{\lambda} \rangle = \sum_{n=0}^{\infty} f_{n+1} \, c_{n+1}^{* (s)} ( z ) 
\label{bargmann1}
\ee
a complex-valued function, and $f_{n+1} \equiv \langle \overline\psi_{n+1} \vert f_{\lambda} \rangle \in \mathbb C$.

\subsection{The Fock-Bargmann spaces}

The realization ${\cal F}$ of a Hilbert space ${\cal H}$ in terms of entire analytic functions \cite{Boa54} is named after Fock \cite{Foc28} and Bargmann \cite{Bar61}. For the harmonic oscillator the space ${\cal F}$ is formed by entire analytic functions of growth $( \frac12, 2)$. Besides, the representation of the ladder operators in ${\cal F}$ is given by $\hat a= 2 \frac{\partial}{\partial \alpha^*}$ and $\hat a^{\dagger} = \alpha^*$. Although this was originally realized for the space of states of the harmonic oscillator, the Fock-Bargmann representation may be related to any coherent state system \cite{Per86}. Next we obtain this for the generalized coherent states of the non-Hermitian oscillators.

Using (\ref{idengral}), the bi-product of $\vert \overline g_{\lambda} \rangle \in \overline{\cal H}_{\lambda}$ with $\vert f_{\lambda} \rangle \in {\cal H}_{\lambda}$ is
\be
\langle \overline g_{\lambda} \vert f_{\lambda} \rangle= \overline g_0^* f_0 + \int \overline g_s^*(z) f_s(z^*) d \sigma_s (z), \quad s={\cal N}, w,
\ee
with measure 
\be
d\sigma_s(z) = \vert c_0^{(s)} (\vert z \vert)  \vert^2 \, d\mu^{(s)} (z).
\label{msigma}
\ee
In particular, from (\ref{prod3}) we know that the bi-norm of $\vert f_{\lambda} \rangle$ is non-negative
\be
\vert \vert \, \vert f_{\lambda} \rangle \vert \vert^2 = \vert f_0 \vert^2 +\int \vert f_s(z^*) \vert^2 d \sigma_s(z) \geq 0.
\label{prev1}
\ee
Therefore, the set of  functions (\ref{bargmann1}) can be equipped with the inner product 
\be
(g,f)_s=\int_{\mathbb C}  \overline g_s^*(z) f_s(z^*) d \sigma_s (z).
\label{inner}
\ee
Besides, as $\vert f_0 \vert^2 \geq 0$, from (\ref{prev1}) we have the non-negative quantity
\be
\vert\vert f_s(z^*) \vert\vert^2 = (f,f)_s= \int_{\mathbb C} \vert f_s(z^*) \vert^2 d \sigma_s(z) \geq 0,
\label{inner2}
\ee
which is useful to introduce the norm of $f_s(z^*)$ as $\vert\vert f_s(z^*) \vert\vert \ = \sqrt{ (f,f)_s }$.

The set of functions $f_s(z^*)$ defined in (\ref{bargmann1}) integrate a vector space ${\cal F}_s$ over $\mathbb C$ with the conventional notions of addition of functions and multiplication by scalars. As indicated above, ${\cal F}_s$ is a normed space with norm $\vert \vert \cdot \vert \vert$ induced by the inner product (\ref{inner}). We assume that ${\cal F}_s$ is complete (the proof is out of the scope of this work), so that it is a Hilbert space which we shall call the {\em Fock-Bargmann} space associated to the coherent states (\ref{csgral}). Consistently, we say that  the Fock-Bargmann representation of  the vector $\vert f_{\lambda} \rangle \in {\cal H}_{\lambda}$ is given by the complex-valued function $f_s(z^*) \in {\cal F}_s$.

To identify some of the fundamental properties of the functions in ${\cal F}_s$ notice that $\vert \langle  \overline \phi^{(s)} (z)  \vert f_{\lambda} \rangle \vert \leq \vert \vert \, \vert f_{\lambda} \rangle \vert \vert$, since the vector $\vert \overline \phi^{(s)} (z) \rangle$ in (\ref{opt1}) is normalized. Then
\be
\vert f_s(z^*) \vert \leq \frac{1}{c_0^{* (s)} (\vert z \vert) }
\vert \vert \, \vert f_{\lambda} \rangle \vert \vert.
\ee
In other words,  provided $\vert \vert \, \vert f_{\lambda} \rangle \vert \vert < \infty$, the asymptotic behavior of $\vert f_s(z^*) \vert$ is bounded from above by the reciprocal of the normalization constant $c_0^{* (s)} (\vert z \vert)$. This last is expressed in terms of the generalized hypergeometric function ${}_pF_q (a_j, b_j; z )$, which in turn satisfies \cite{Ros96}
\be
\left\vert \sqrt{ {}_pF_q (a_j, b_j; z^{\ell} )} \right\vert = \sqrt{ \vert {}_pF_q (a_j, b_j; z^{\ell} ) \vert } \leq \exp \left( \tfrac12  \sigma_{\ell} r^{\rho_{\ell}} \right), \quad z =re^{i\phi}.
\ee
That is, the reciprocal of $c_0^{* (s)} (\vert z \vert)$ is an entire (exponential) analytic function of order $\rho_{\ell}$ and type $\frac12 \sigma_{\ell} $, with $\sigma_{\ell}= \ell/\rho_{\ell}=1+q-p$. In conclusion, ${\cal F}_s$ is the space formed by the entire analytic functions $f_s(z^*)$ of growth $( \frac12 \sigma_{\ell}, \rho_{\ell})$ that satisfy (\ref{inner}) and (\ref{inner2}). In particular, from (\ref{ec3}) and (\ref{constdist}) we realize that $f_{\cal N} (z^*)$ and $f_w (z^*)$ are entire analytic functions of growth $( \frac32, \frac23)$ and $( \frac12, 2)$, respectively. Quite interestingly, the Fock-Bargmann space ${\cal F}_w$ coincides with ${\cal F}$.

\subsubsection{Canonical operators in ${\cal F}_{\cal N}$}

Considering that $\vert \overline\phi^{({\cal N})} (z)\rangle_U$ is eigenvector of $\overline{\cal A}$ with complex eigenvalue $z$, and using $\overline{\cal A}^{\dagger} ={\cal A}^+$, we have
\be
{}_U\langle \overline\phi^{({\cal N})} \vert \overline{\cal A}^{\dagger} \vert f_{\lambda} \rangle :=  {\cal A}^+_{op} f_{\cal N}(z^*) =z^*  f_{\cal N}(z^*).
\ee
That is, the multiplication by $z^*$ in ${\cal F}_{\cal N}$ is achieved by the action of ${\cal A}^+_{op}$ on $f_{\cal N}(z^*)$. On the other hand, the straightforward calculation shows that ${\cal A}$ is represented in ${\cal F}_{\cal N}$ by the third-order differential operator
\be
{\cal A}_{op}  f_{\cal N} (z^*) = \frac{1}{\sqrt 2} \left[ 4z^{*2} \frac{\partial^3}{\partial z^{*3}} +4(3-\epsilon) z^* \frac{\partial^2}{\partial z^{*2}} + (1-\epsilon)(3-\epsilon) \frac{\partial}{\partial z^*}
\right] f_{\cal N} (z^*),
\label{fockop1}
\ee
where
\be
\frac{\partial}{\partial z^*} f_{\cal N} (z^*) = \sum_{n=0}^{\infty}  f_{n+2}  \left[ \frac{\Gamma \left( \frac{1- \epsilon}{2} \right)  \Gamma \left(\frac{3- \epsilon}{2} \right) }{ n! \, \Gamma \left( n + \frac{3-\epsilon}{2} \right) \Gamma \left( n + \frac{5-\epsilon}{2} \right) } \right]^{1/2}  \sqrt{ \frac{n+1}{8} } \left( \frac{z^*}{\sqrt{8} } \right)^n.
\label{fockop2}
\ee
Therefore, ${\cal A}_{op} \neq \frac{\partial}{\partial z^*}$ for any $\epsilon <E_0=1$. In other words, ${\cal A}_{op}$ is not the canonical conjugate of ${\cal A}^+_{op}$ in ${\cal F}_{\cal N}$. The latter is a consequence of the quadratic algebra (\ref{Acomm1})-(\ref{Acomm2}), which is satisfied by ${\cal A}$ and ${\cal A}^+$, and the transformations (\ref{htrans}). Looking for the simplest operator ${\cal A}_{\partial} \in \mbox{Aut}({\cal H}_{\lambda})$ that is represented by the $z^*$-derivative in ${\cal F}_{\cal N}$, one can show that it is such that its conjugate $\overline{\cal A}^+_{\partial}$ operates as follows
\be
\overline{\cal A}^+_{\partial} \vert \overline\psi_{n+1} \rangle =  \left[ \frac{ 2(n+1)}{ (2n+1-\epsilon)(2n+3-\epsilon) } \right]^{1/2}\vert \overline\psi_{n+2} \rangle, \quad n\geq 0.
\ee
Thus, ${\cal A}_{\partial,op} = 2\frac{\partial}{\partial z^*}$. We see that it is ${\cal A}_{op}^{\dagger}$ and ${\cal A}_{\partial,op}$, and not ${\cal A}_{op}^{\dagger}$ together with ${\cal A}_{op}$, the pair of operators in ${\cal F}_{\cal N}$ that behave as $\hat a^{\dagger}$ and $\hat a$ in ${\cal F}$.

\subsubsection{Canonical operators in ${\cal F}_w$}

Following the steps of the previous section one can verify that 
\be
 {\cal C}^+_{w,op} f_w (z^*)  := {}_e\langle \overline\phi^{(w)} \vert \overline{\cal C}_w^{\dagger} \vert f_{\lambda} \rangle =z^*  f_w (z^*),
\ee
and 
\be
{\cal C}_{w,op} f_w (z^*) = 2 \frac{\partial}{\partial z^*} f_w (z^*) + 2 \frac{(w-1)}{z^*} \left( f_w(z^*) -f_1 \right),
\ee
with
\be
\frac{\partial}{\partial z^*} f_w (z^*) = \frac{1}{\sqrt 2} \sum_{n=0}^{\infty} f_{n+2}  \left[ \frac{ \Gamma(w) }{ \Gamma(w+n+1)} \right]^{1/2} (n+1) \left( \frac{z^*}{\sqrt 2}
\right)^n.
\label{fockop3}
\ee
Thus, ${\cal C}^+_{w,op}$ and ${\cal C}_{w,op}$ behave in ${\cal F}_w$ just as $\hat a^{\dagger}$ and $\hat a$ in ${\cal F}$ for $w=1$. In contrast with ${\cal A}_{op}^{\dagger}$ and ${\cal A}_{op}$, the closest behavior of ${\cal C}^+_{w,op}$ and ${\cal C}_{w,op}$ with the canonical operators of the harmonic oscillator is due to the fact that the distorted Heisenberg algebra (\ref{rcom2}), (\ref{dist2}), is linear.

\subsection{The $P$-representation}

The $P$-representation was introduced by Glauber \cite{Gla63} and Sudarshan \cite{Sud63} for the states $\hat \rho$ of the harmonic oscillator. This is defined such that $\hat \rho$ can be written in the basis of the coherent states of the harmonic oscillator as the continuous `diagonal' matrix
\be
\hat \rho= \frac{1}{2\pi} \int_{\mathbb C} d^2\alpha \, P(\alpha) \, \vert \alpha \rangle \langle \alpha \vert.
\label{prep}
\ee
If the $P$-function in (\ref{prep}) behaves as a probability, then $\hat \rho$ can be interpreted as an statistical ensemble (i.e., a mixed state) of the pure states $\vert \alpha \rangle \langle \alpha \vert$. Otherwise, the system represented by $\hat \rho$ ``will have no classical analog'' \cite{Gla07}. In order to include the $P$-representation of any coherent state of the harmonic oscillator $\hat \rho_z = \vert z \rangle \langle z \vert$, the function $P(\alpha)$ is permitted to be as singular as the delta function $\delta^{(2)}(z-\alpha) =\delta (\mbox{Re}(z-\alpha)) \delta (\mbox{Im}(z-\alpha))$. In this context, the number eigenstates $\vert n \rangle$ are nonclassical for $n>1$ because their $P$-functions are determined by the derivatives of the delta function. In turn, the vacuum $\vert 0 \rangle$ is classical because it is a coherent state with complex eigenvalue equal to zero. 

In the present case, given any state represented by the density operator $\rho \in \mbox{Aut} ({\cal H}_{\lambda})$, we want to use the identity operator (\ref{idengral}) to find the $P$-representation
\be
\rho= P_0 \vert \psi_0 \rangle \langle \overline\psi_0 \vert  + \int d\mu^{(s)} (z) \, P^{(s)}(z,z^*) \,  \vert \phi^{(s)} (z) \rangle \langle \overline \phi^{(s)} (z) \vert, \quad s={\cal N}, w.
\label{P1}
\ee
As $\vert \psi_0 \rangle$ is bi-orthogonal to all the $\vert \phi^{(s)} (z) \rangle$, we immediately have
\be
P_0 = \langle \overline\psi_0 \vert \rho \vert \psi_0 \rangle.
\label{P0}
\ee
To determine $P^{(s)}(z,z^*)$ let us consider an arbitrary state $\vert \beta \rangle \neq \vert \psi_0 \rangle$ written as
\be
\vert \beta \rangle = \sum_{n=0}^{\infty} \beta_n \vert \psi_{n+1} \rangle, \quad \beta_n \in \mathbb C.
\label{P2}
\ee
Then, the function we are looking for is such that the following expression is true
\be
\langle -\overline\beta \vert \rho \vert \beta \rangle = \int d\sigma^{(s)}(z)  P^{(s)}(z,z^*)  \langle -\overline\beta \vert \phi^{(s)} (z) \rangle_U {}_U\langle \overline \phi^{(s)} (z) \vert \beta \rangle,
\label{P3}
\ee
where we have used (\ref{csgral}) and (\ref{msigma}). As the vector $\vert \beta \rangle$ is arbitrary, to proceed we have two options:  (i) To take $\vert \beta \rangle$ such that
\be
{}_U\langle \overline \phi^{(s)} (z) \vert \beta \rangle = \exp(\beta z^*).
\label{P4}
\ee
In this case the integral (\ref{P3}) acquires the form
\be
\langle -\overline\beta \vert \rho \vert \beta \rangle = \int d\sigma^{(s)}(z)  P^{(s)}(z,z^*) e^{\beta z^*-\beta^* z}.
\label{P5}
\ee
Therefore, $P^{(s)} (z,z^*)$ is obtained from the two-dimensional inverse Fourier transform of this last expression. (ii) To make $\vert \beta \rangle = \vert \phi^{(s)} (y) \rangle$ and, after using the reproducing kernel $K_s(y^*,z)$, to solve (\ref{P3}) for $P^{(s)}(z,z^*)$. Both of the above options lead to the same result, as it can be easily verified.

\subsubsection{The basis of natural coherent states}

For the natural coherent states introduced in Sec.~\ref{natCS}, the vector $\vert \beta \rangle$ that satisfies (\ref{P4}) is properly constructed by taking
\be
\beta_n = \left[ \frac{ \Gamma(n+ \frac{1-\epsilon}{2}) \Gamma (n+ \frac{3-\epsilon}{2})
}{n! \, \Gamma(\frac{1-\epsilon}{2} ) \Gamma( \frac{3-\epsilon}{2}) }\right]^{1/2} ({\sqrt 8} \beta)^n, \quad \beta \in \mathbb C, \quad n\geq 0,
\label{P6}
\ee
as the coefficients of the superposition (\ref{P2}). Then, using (\ref{ec3}), (\ref{mu}) and (\ref{msigma}), the inverse transform of (\ref{P5}) gives
\be
P^{({\cal N})}(z,z^*)= \frac{ \Gamma(\frac{1-\epsilon}{2} ) \Gamma( \frac{3-\epsilon}{2}) }{ \pi^2 h^{({\cal N})} (r^2) } \int d^2 \beta \langle -\overline\beta \vert \rho \vert \beta \rangle \, e^{\beta^* z-\beta z^*}.
\label{P7}
\ee
As an immediate application consider the energy eigenstate $\rho_{n+1}= \vert \psi_{n+1} \rangle \langle \overline \psi_{n+1} \vert$, then
\be
P^{({\cal N})}_{n+1} (z,z^*)= 8^n \frac{ \Gamma(n+ \frac{1-\epsilon}{2}) \Gamma (n+ \frac{3-\epsilon}{2})
}{ n! \, h^{({\cal N})} (r^2) } \frac{ \partial^{2n}}{ \partial z^n \partial z^{*n} } \delta^{(2)}(z), \quad n\geq0.
\label{P8}
\ee
That is, the $P$-function of the first excited energy eigenstate $\vert \psi_1 \rangle$ is as singular as the delta function $\delta^{(2)}(z)$. Now, from (\ref{P0}) we see that the state $\vert \psi_0 \rangle$ is such that $P_0=1$. In turn, the $P$-function of the higher excited states $\vert \psi_{n \geq 2} \rangle$ corresponds to the derivatives of $\delta^{(2)}(z)$. On the other hand, for any natural coherent state $\rho(\alpha) = \vert \phi^{({\cal N})} (\alpha) \rangle \langle \overline \phi^{({\cal N})} (\alpha) \vert$ we have 
\be
P^{({\cal N})}_{\alpha} (z,z^*)= \frac{ \Gamma(\frac{1-\epsilon}{2} ) \Gamma( \frac{3-\epsilon}{2}) }{ h^{({\cal N})} (r^2) {}_0F_2( \frac{1-\epsilon}{2}, \frac{3-\epsilon}{2}; \frac{\vert \alpha \vert^2}{8} )} \delta^{(2)} (z-\alpha),
\label{P9}
\ee
as expected. Besides, comparing (\ref{P9}) with (\ref{P8}) we realize that $\vert \phi^{({\cal N})} (\alpha) \rangle$ is a displaced version of the fiducial state $\vert \psi_1 \rangle$.

\subsubsection{The basis of distorted coherent states}

For the distorted coherent states introduced in Sec.~\ref{distCS} one can verify that 
\be
P^{(w)}_{n+1}(z,z^*)= \frac{ \pi 2^n \Gamma(w+n) }{ (n!)^2 r^{2(w-1)} h^{(w)}(r^2) } \frac{ \partial^{2n} }{ \partial z^n \partial z^{*n} } \delta^{(2)}(z), \quad n\geq 0,
\label{P10}
\ee
is the $P$-function associated to the energy eigenstates $\vert \psi_{n+1} \rangle$. That is, as in the previous case, the energy states $\vert \psi_0 \rangle$ and $\vert \psi_1 \rangle$ are $P$-represented by $P_0=1$ and $P_1(z,z^*)= \delta^{(2)}(z)$ respectively. In turn, for the distorted coherent state $\vert \phi^{(w)}  (\alpha)\rangle$ we have

\be
P^{(w)}_{\alpha}(z,z^*)= \frac{ \pi \Gamma(w) }{{}_1F_1(1,w; \frac{\vert \alpha \vert^2}{2} ) \,r^{2(w-1)} h^{(w)}(r^2) } \delta^{(2)}(z-\alpha).
\label{P11}
\ee
Clearly, for $\alpha=0$ the equation (\ref{P11}) reproduces (\ref{P10}) with $n=0$. That is, $\vert \phi^{(w)}  (\alpha)\rangle$ is a displaced version of the fiducial state $\vert \psi_1 \rangle$.

\section{Concrete models and nonlinearity}
\label{equidistant}

\subsection{Equidistant spectrum}
\label{experiment}

For $\epsilon = -1$ in (\ref{alpha1})-(\ref{alpha2}), the complex-valued potential (\ref{pot2a}) acquires the form
\be
\left. V_{\lambda}(x) \right\vert_{\epsilon=-1} = x^2 -2 -2\frac{d}{dx} \left( \frac{2a \mbox{Erf}(x) + b -i\sqrt{\pi} \lambda}{\sqrt{\pi} \alpha^2(x)} \right).
\label{potosc}
\ee
The energy spectrum is now equidistant and shifted in two units with respect to the harmonic oscillator energies $E^{(\lambda)}_n =2n-1$, $n=0,1,2, \ldots$ That is, besides the harmonic oscillator Hamiltonian $H=-\frac{d^2}{dx^2} +x^2$, either of the non-Hermitian Hamiltonians $\left. H_{\lambda} \right\vert_{\epsilon =-1}+2$ or $\left. \overline H_{\lambda} \right\vert_{\epsilon =-1}+2$ can be used to represent the observable associated with a measurement of energy which gives any of the real numbers $E_n=2n+1$ as result. The predictive character of such a model is sustained by the bi-orthogonal structure discussed in the previous sections. 
 
\begin{figure}[htb]
\centering

\centering
\subfigure[$\epsilon=-1$]{\includegraphics[width=0.3\textwidth]{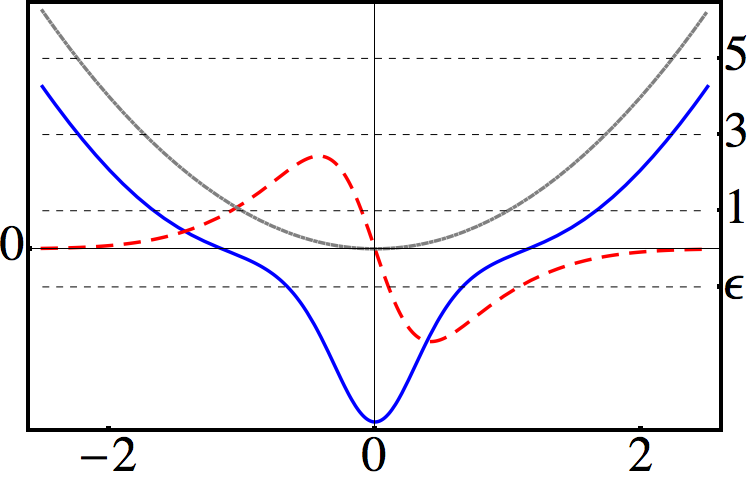}} 
\hspace{3ex}
\subfigure[$\epsilon=-1$]{\includegraphics[width=0.3\textwidth]{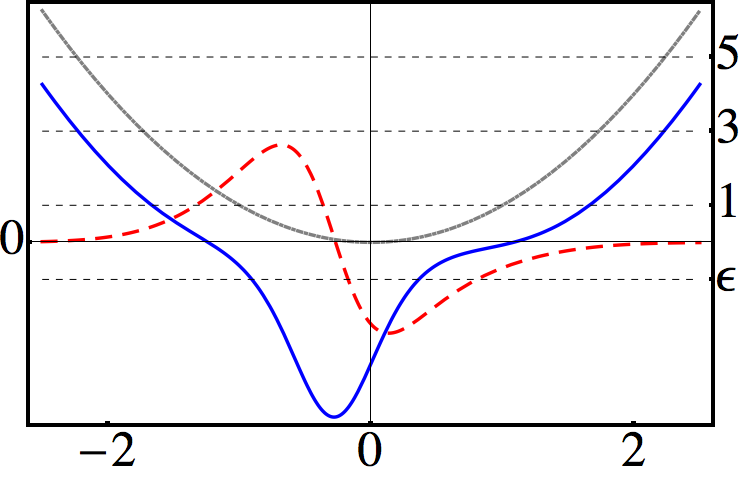}}

\caption{\footnotesize 
(Color online) The real (solid-blue) and imaginary (dashed-red) parts of the complex-valued potential (\ref{potosc}) for $a=\frac{\pi}{4}$, $c=1$, with 
(a) $b=0$, $\lambda=\frac{\sqrt \pi}{2}$, and
(b) $b=\frac{\sqrt \pi}{2}$, $\lambda=\frac{\sqrt{3\pi} }{4}$. The oscillator potential (dotted-gray) is included as a reference. The horizontal dotted lines correspond to the first allowed energies. In (a) the potential is ${\cal PT}$-symmetric. 
}
\label{f_eq}
\end{figure}

In Fig.~\ref{f_eq} we show the behavior of the non-Hermitian oscillators (\ref{potosc}) for two different sets of parameters such that the condition of zero total area (\ref{zero}) is fulfilled. The potential depicted in Fig~\ref{f_eq}(a) is ${\cal PT}$-symmetric.

\subsection{Hermitian oscillator-like Hamiltonians}

For $\lambda =0$ the operator $H_{\lambda=0} = \overline H_{\lambda=0} = -\frac{d^2}{dx^2} +V_{\lambda=0}(x)$ is self-adjoint but it is different from the harmonic oscillator Hamiltonian. Some examples of $V_{\lambda=0}(x)$ are depicted as solid-blue curves in Figs.~\ref{f_pot} and \ref{f_eq}; the difference with the harmonic oscillator potential  (dotted-gray curves in the figures) is notable. Besides, with $\lambda =0$ the eigenvectors $\vert \psi_n \rangle$ and their concomitants $\vert \overline \psi_n \rangle$ collapse to the same complete set of orthonormal vectors $\vert \psi_n \rangle_{\lambda=0} =  \vert \overline \psi_n \rangle_{\lambda=0}$. Thus, while the quadratic polynomial and distorted Heisenberg algebras are preserved, the representation space defined by the bi-orthogonal system $\{ \vert \overline \psi_n \rangle, \vert \psi_m \rangle \}_{n,m \geq 0}$ converges to the complete orthonormal set of vectors $\{ \vert \psi_n \rangle_{\lambda=0} \}_{n \geq 0}$. 

We would like to emphasize that all the results derived in the previous sections are maintained without changes in the Hermitian picture defined by $\lambda =0$. The only difference lies on the fact that the intertwining relationships, (\ref{inter1}) and (\ref{inter2}), are now referred to a pair and not to a triad of systems (cf. Fig~\ref{f_spectra}). Such systems are represented by $H_{\lambda=0}$ and $H$, and have a complete set of eigenvectors by their own.

\subsubsection{Abraham-Moses-Mielnik (AMM) oscillators}

The combination of the parameters used in the two previous sections gives rise to the self-adjoint Hamiltonian $H_{\gamma} = H_{\lambda=0, \epsilon =-1} \equiv-\frac{d^2}{dx^2} +V(x;\gamma)$, with 
\be
V(x; \gamma)= x^2-2-2\frac{d}{dx} M_{\gamma}(x), \quad 
M_{\gamma}(x) = \frac{e^{-x^2}}{\gamma + \int^x e^{-y^2} dy}, \quad \gamma = \sqrt{ \frac{\pi c}{4a}} \in \mathbb R
\label{potg}
\ee
a real-valued potential which was found independently by Abraham and Moses through the systematic use of the Gelfand-Levitan equation \cite{Abr80}, and by Mielnik as an application of his generalized factorization method \cite{Mie84}. The AMM oscillators $H_{\gamma}$ are intertwined with the harmonic oscillator $H$ via (\ref{inter1}) and (\ref{inter2}) with
\be
\left. A \right\vert_{\lambda=0, \epsilon=-1} = \frac{d}{dx} + x + M_{\gamma}(x) ,\quad  
\left. B \right\vert_{\lambda=0, \epsilon=-1} = - \frac{d}{dx} + x + M_{\gamma}(x),
\label{newab}
\ee
where a global phase has been dropped out.
 
The natural coherent states (\ref{ec1})-(\ref{ec3}) acquire the form
\be
\vert \phi^{({\cal N})} (z)  \rangle_{\lambda=0, \epsilon=-1}  = \frac{1}{ \sqrt{ {}_0 F_2 (1,2, \vert z \vert^2/8) } } \sum_{n=0}^{+\infty} \frac{1}{n! \sqrt{ \Gamma(n+2)} } \left( \frac{z}{\sqrt 8} \right)^n \vert \psi_{n+1} \rangle_{\lambda=0}.
\label{csnat}
\ee
The vectors (\ref{csnat}) coincide (up to a factor in the complex number $z$) with the generalized coherent states reported in \cite{Fer94} for the AMM oscillators. Moreover, the $P$-representations (\ref{P8}) and (\ref{P9}) are unchanged. On the other hand, as the distorted coherent states (\ref{distcoh})-(\ref{constdist}) do not depend on the ground energy $\epsilon < E_0=1$, they preserve their form
\be
\vert \phi^{(w)} (z)  \rangle_{\lambda=0} =\sqrt{  \frac{ \Gamma(w) }{ {}_1F_1 \left( 1,w; \vert z \vert^2/2 \right) } }  \sum_{n=0}^{+\infty}  \frac{ (z/\sqrt{2})^n }{ \sqrt{ \Gamma(w+n)} } 
\vert \psi_{n+1} \rangle_{\lambda=0}.
\label{distinv}
\ee
Up to a factor in $z$, these last vectors coincide with the generalized coherent states introduced in \cite{Fer95,Ros96} for the AMM oscillators. The $P$-representations (\ref{P10}) and (\ref{P11}) are also preserved.

\subsection{Quantum oscillator limit}
 
We can go a step further by cancelling the term with $M_{\gamma}(x)$ in (\ref{potg}). That is, 
\be
\lim_{\gamma \rightarrow +\infty} V(x; \gamma)= x^2-2.
\ee
In other words, the Hamiltonian $H_{\gamma}+2$ converges to $H$ at $\gamma \rightarrow +\infty$. Consistently, the wave-functions $\psi_n(x) := \langle x \vert \psi_n \rangle$ go to the number eigenfunctions $\varphi_n(x):= \langle x \vert n \rangle$ under the rule
\be
\lim_{\gamma \rightarrow \infty} \left. \psi_n(x) \right\vert_{\lambda=0, \epsilon=-1} = \varphi_n(x), \quad n\geq 0.
\label{states}
\ee
In turn, the intertwining operators (\ref{newab}) become the boson ladder operators
\be
\lim_{\gamma \rightarrow \infty} \left. A \right\vert_{\lambda=0, \epsilon=-1} =  \hat a, \quad \lim_{\gamma \rightarrow \infty} \left. B \right\vert_{\lambda=0, \epsilon=-1} =  \hat a^{\dagger}.
\ee
The latter result justifies the structure of the vectors (\ref{pad2}) since they were obtained from the the action of $B$ on the coherent state $\vert \alpha \rangle \in {\cal H}$ for $\epsilon =-1$. If now we make $\lambda=0$, it is clear that the vectors $\left. \vert \alpha^{(\lambda =0)} \rangle \right\vert_{\epsilon =-1}$ in (\ref{pad2}) coincide with the conventional one-photon added coherent states \cite{Aga91,Siv99}.

Notice however that the generators of the quadratic polynomial Heisenberg algebra converge to the $f$-boson operators
\be
{\cal A}_{osc} = (2 \hat N ) \hat a \equiv \hat a_N, \qquad {\cal A}^{\dagger}_{osc} =  \hat a^{\dagger} (2 \hat N)  \equiv \hat a_N^{\dagger},
\label{aosc}
\ee
while the distorted ladder operators become
\be
{\cal C}_{w,osc} = \hat a_{f_w}^{\dagger} \hat a^2 , \qquad {\cal C}_{w,osc}^{\dagger} = \left( \hat a^{\dagger} \right)^2 \hat a_{f_w}.
\label{cosc}
\ee
The above operators represent nonlinear interactions for the harmonic oscillator. The action of the pair defined in (\ref{aosc}) produces transitions between the number eigenvectors $\vert n \rangle$ that are mediated by the number operator. Considering a single-mode photon field described by the harmonic oscillator, the operators (\ref{aosc}) represent intensity dependent interactions with a two-level atom in the well-known Jaynes-Cummings model \cite{Jay63} (see details in \cite{Enr17}). In the same picture, the pair of operators (\ref{cosc}) involve three photons in the process. Namely, ${\cal C}_{w,osc}^{\dagger}$ implies the annihilation of one $f_w$-photon at the time that two new photons are created. The Hermitian-conjugate ${\cal C}_{w,osc}$ operates in reverse order and is useful to get self-adjoint expressions which could represent the involved photon-fields. The process resembles the {\em spontaneous parametric down conversion} that occurs, for example, when a nonlinear crystal is illuminated by the appropriate light (see, e.g. \cite{Pro14} and references quoted therein); this multi-photon phenomenon is currently observed in the laboratory \cite{Pro10,Cal16}. As we can see, the operators (\ref{aosc})-(\ref{cosc}) would be useful for modeling nonlinear phenomena in quantum optics (see e.g. \cite{Mat96,Roy00a,Roy00b,Dod17}).

\subsection{Nonlinear coherent states of the harmonic oscillator}

In this section we discuss some of the properties of the generalized coherent states associated with the nonlinear operators (\ref{aosc})-(\ref{cosc}) discussed above.

The nonlinear coherent states $\vert \phi_{osc}^{({\cal N})} (z)  \rangle$ associated with the quadratic polynomial operators (\ref{aosc}) have the same structure as the vectors (\ref{csnat}), with $\vert \psi_{n+1} \rangle_{\lambda=0}$ substituted by $\vert n+1 \rangle$. The $P$-representation for the energy eigenvectors of the oscillator is obtained from (\ref{P8}) and gives
\be
\left. P^{({\cal N})}_{n+1} (z,z^*) \right\vert_{osc}= 8^n  \frac{ \Gamma(n+2) }{ h^{({\cal N})} (r^2)} \frac{\partial^{2n}} {\partial z^n \partial z^{*n}} \delta^{(2)}(z), \quad n\geq0,
\label{csnatP1}
\ee
with $P_0=1$ for the ground state $\vert 0 \rangle$. That is, in the basis of the nonlinear natural coherent states, the first excited energy eigenstate $\vert 1 \rangle$ of the harmonic oscillator is $P$-represented by the delta distribution. Such a striking result is a consequence of the nonlinearity of the operators (\ref{aosc}), which is inherited from the quadratic polynomial structure discussed in the previous sections.

Now, from (\ref{P9}) we see that the nonlinear natural coherent states $\vert \phi_{osc}^{({\cal N})} (z)  \rangle$ are also $P$-represented by the delta distribution 
\be
\left. P^{({\cal N})}_{\alpha} (z,z^*) \right\vert_{osc}= \left[ h^{({\cal N})}(r^2) {}_0 F_2 (1,2,\vert z \vert^2/8) \right]^{-1} \delta^{(2)} (z-\alpha),
\label{csnatP2}
\ee
and that they are displaced versions of the fiducial state $\vert 1 \rangle$.

On the other hand, the nonlinear distorted coherent states $\vert \phi^{(w)}_{osc} (z)  \rangle$ are obtained from (\ref{distinv}), with $\vert \psi_{n+1} \rangle_{\lambda=0}$ changed for $\vert n+1 \rangle$. The $P$-representation for the number states $\vert n \rangle$ has been already given in (\ref{P10}). Notably, also in the basis $\vert \phi^{(w)}_{osc} (z)  \rangle$ we find that $\vert 0 \rangle$ and $\vert 1 \rangle$ are $P$-represented by $P_0=1$ and $P_1 (z,z^*)=\delta^{(2)}(z)$. Of course, the nonlinear distorted coherent states (\ref{distinv}) are  displaced versions of the fiducial state $\vert 1 \rangle$ since their $P$-representation (\ref{P11}) is also proportional to the $\delta$-function. 

\section{Conclusions}
\label{conclu}

We have constructed a bi-orthogonal system for a series of Hamiltonians $H_{\lambda}$ that are not self-adjoint but have the spectrum of the harmonic oscillator plus an additional eigenvalue located below the ground energy of the latter. The operators $H_{\lambda}$ are Darboux transformations of the conventional oscillator Hamiltonian $H$ such that their potentials are complex-valued. The bi-orthogonality between the states of $H_{\lambda}$ and those of its Hermitian-conjugate $H_{\lambda}^{\dagger}$ provides a mathematical structure that ensures the fulfilling of the superposition principle. In this form, the challenge of constructing generalized coherent states for the non-Hermitian oscillators represented by $H_{\lambda}$ and $H_{\lambda}^{\dagger}$ is faced in much the same form as in the Hermitian approaches. Two different algebras have been found for these non-Hermitian oscillators. One of them is defined by the ground energy of the system and is a quadratic polynomial variation of the Heisenberg algebra. The other one is lineal and depends on a non-negative parameter in such a way that the Heisenberg algebra is fulfilled in definite subspaces of the space of states. This is called distorted Heisenberg algebra and is attainable by all the non-Hermitian oscillators studied here, no matter the position of the ground state energy. The generalized coherent states constructed from these algebras were used to obtain the $P$-representation of the eigenvectors of the non-Hermitian oscillators. It is found that the first excited energy eigenstate $\vert \psi_1 \rangle$ is $P$-represented by a delta distribution. Such a striking behavior is not shared with the harmonic oscillator since all the excited states of the latter are $P$-represented by the derivatives of the $\delta$-function. Besides, it is $\vert \psi_1 \rangle$ which serves as fiducial state for the generalized coherent states. This is because the annihilation operators in both of the algebras  annihilate the ground state $\vert \psi_0 \rangle$ as well as $\vert \psi_1 \rangle$. Thus, the complex eigenvalue $z=0$ is twice degenerate for all the generalized coherent states of $H_{\lambda}$ and $H_{\lambda}^{\dagger}$ that are constructed as eigenvectors of the annihilation operators. Moreover, in $P$-representation such states are displaced versions of $\vert \psi_1 \rangle$.

We can take full advantage of the above property to construct additional sets of coherent states. Consider for example the distorted Heisenberg algebra developed in Sec.~\ref{distorted}. As ${\cal C}_w \vert \psi_1 \rangle ={\cal C}_w \vert \psi_0 \rangle = {\cal C}_w^+ \vert \psi_0 \rangle =\vert  \varnothing \rangle$, we have
\be
D_w(z) \vert \psi_0 \rangle = \vert \psi_0 \rangle, \quad D_w(z) \vert \psi_1 \rangle = e^{z {\cal C}_w^+} \vert \psi_1 \rangle \propto \vert \phi^{(wd)} (z,w) \rangle,
\ee 
with
\be
D_w(z) = e^{z {\cal C}_w^+}  e^{-z^* {\cal C}_w}
\label{Dop}
\ee 
an operator that leaves invariant the ground energy eigenstate $\vert \psi_0 \rangle$ but `displaces'  $\vert \psi_1 \rangle$ to the (normalized) state
\be
\vert \phi^{(wd)} (z,w) \rangle = \frac{1}{ \sqrt{ {}_1F_1 (w,1, 2 \vert z \vert^2)}}  \sum_{n=0}^{\infty} \left[ \frac{ \Gamma(w+n) }{ \Gamma(w) } \right]^{1/2} \frac{(\sqrt{2} z)^n }{n!} \vert \psi_{n+1} \rangle.
\ee
This last is also a generalized coherent state associated with the non-Hermitian oscillators studied in the present work, with properties that are quite similar to those of the other coherent states discussed here. However, the operator $D_w(z)$ deserves attention since the commutator between ${\cal C}_w$ and ${\cal C}_w^+$ is nontrivial and it is not easy to guess a disentangling formula for (\ref{Dop}). Work in this direction is in progress.

To conclude we would like to emphasize that besides the harmonic oscillator Hamiltonian $H$, either of the non-Hermitian Hamiltonians $H_{\lambda}$ and $H_{\lambda}^{\dagger}$ (for the appropriate parameters see Sec.~\ref{experiment}) can be used to represent the observable associated with a measurement of energy which gives any of the real numbers $E_n = 2n+1$ as result. The predictive character of such a model is sustained by the bi-orthogonal structure discussed along this paper. Is it then permissible to assign uniquely the self-adjoint operator $H$ to the observable of energy in such a class of measurement? We hope that the present work shed new light on the subject. 

\section*{Acknowledgment}

We acknowledge the financial support from the Spanish MINECO (Project MTM2014-57129-C2-1-P) and Junta de Castilla y Le\'on (VA057U16). K.~Zelaya acknowledges the support of CONACyT, scholarship 45454. ORO is in debt with Anna Okopi\'nska and Faruk Gungor for their valuable comments.


\end{document}